\documentclass{article}

\usepackage{arxiv}
\usepackage{authblk}

\usepackage[utf8]{inputenc} 
\usepackage[T1]{fontenc}    
\usepackage[hidelinks]{hyperref}       
\usepackage{url}            
\usepackage{booktabs}       
\usepackage{amsfonts}       
\usepackage{nicefrac}       
\usepackage{microtype}      

\usepackage{amsmath}
\usepackage{cleveref}       
\usepackage{lipsum}         
\usepackage{graphicx}
\usepackage{natbib}
\usepackage{doi}

\usepackage{times}
\usepackage{subcaption}

\usepackage{natbib}

\usepackage[nobiblatex]{xurl}

\usepackage{enumerate}

\usepackage{rotating}
\usepackage{threeparttable}
\usepackage{multirow}
\usepackage{placeins}
\usepackage{diagbox}
\usepackage{algpseudocode,algorithm,algorithmicx}

\def\max_cards{N}
\def\customer_cards{n}
\def\balance{B}
\def\credlimit{L}

\def\total_exp{X}
\def\credexp{E}
\def\income{I}
\def\score{S}
\def\expected_cost{C}

\def\utilisation_component{U}
\def\offer_component{O}

\def\appconst{\rho}

\def\apptend{\lambda_{\text{tend}}}
\def\appminprob{\lambda_{\text{0}}}

\def\better_offer_weighting{\rho_{\text{o}}}
\def\applimitweighting{\lambda}

\def\apprationality{\lambda_{\text{r}}}
\def\miss_pay_prob{P_{\text{miss}}}
\def\usage_tendency{\theta_{\text{tend}}}
\def\cred_lim_weighting{\phi_{\text{\credlimit}}}
\def\repaytend{\eta_{\text{tend}}}
\def\repayconst{\eta}

\def\interestweighting{\mu_{\text{i}}}
\def\balanceweighting{\mu_{\text{B}}}

\def\is_real_number{\in \mathbb{R}}

\title{Agent-based Modelling of Credit Card Promotions}
\date{}

\author[1]{C. B. Hamill}
\author[1,2]{R. Khraishi}
\author[1]{S. Gherghel}
\author[1]{J. Lawrence}
\author[1,$\star$]{S. Mercuri}
\author[2]{R. Okhrati}
\author[1]{G. A. Cowan}
\affil[1] {%
	Data Science \& Innovation\\
	NatWest Group\\
	United Kingdom$\textsuperscript{\dag}$}
\affil[2] {%
	University College London\\
	United Kingdom$\textsuperscript{\ddag}$}

\renewcommand{\headeright}{}
\renewcommand{\undertitle}{}
\renewcommand{\shorttitle}{Agent-based modelling of credit card promotions}

\begin{document}

\maketitle
\begingroup\def\thefootnote{$\star$}\footnotetext{No longer at institute\textsuperscript{1}}\endgroup
\begingroup\def\thefootnote{$\dag$}\footnotetext{Correspondence to: conor.hamill@natwest.com}\endgroup
\begingroup\def\thefootnote{$\ddag$}\footnotetext{Correspondence to: raad.khraishi@ucl.ac.uk}\endgroup

\begin{abstract}
Interest-free promotions are a prevalent strategy employed by credit card lenders to attract new customers, yet the research exploring their effects on both consumers and lenders remains relatively sparse.
The process of selecting an optimal promotion strategy is intricate, involving the determination of an interest-free period duration and promotion-availability window, all within the context of competing offers, fluctuating market dynamics, and complex consumer behaviour.
In this paper, we introduce an agent-based model that facilitates the exploration of various credit card promotions under diverse market scenarios.
Our approach, distinct from previous agent-based models, concentrates on optimising promotion strategies and is calibrated using benchmarks from the UK credit card market from 2019 to 2020, with agent properties derived from historical distributions of the UK population from roughly the same period.
We validate our model against stylised facts and time-series data, thereby demonstrating the value of this technique for investigating pricing strategies and understanding credit card customer behaviour.
Our experiments reveal that, in the absence of competitor promotions, lender profit is maximised by an interest-free duration of approximately 12 months while market share is maximised by offering the longest duration possible.
When competitors do not offer promotions, extended promotion availability windows yield maximum profit for lenders while also maximising market share.
In the context of concurrent interest-free promotions, we identify that the optimal lender strategy entails offering a more competitive interest-free period and a rapid response to competing promotional offers. 
Notably, a delay of three months in responding to a rival promotion corresponds to a 2.4\% relative decline in income.
\end{abstract} 


\section{Introduction}
\label{introduction_label}

Credit card promotions are a widespread tool for lenders to attract new customers, with a large number of offers provided by lenders.
Credit cards remain a popular choice for consumers, due to their enduring appeal as a convenient source of credit, with £17.3bn spent using credit cards in February 2023 in the UK.\footnote{\href{https://ukfinance.org.uk/data-and-research/data/card-spending}{https://www.ukfinance.org.uk}}
Despite the significant profits that credit cards can generate, and their ubiquity as a payment method, research on the effects of certain aspects of different credit card pricing strategies, including promotions offered, remains limited.

Lenders offer various promotions on lending products as part of their pricing strategies, with the aim of attracting new customers and potentially increasing long-term income.
These promotions include introductory promotional rates or ``teaser rates'', where new adopters of the credit card will pay zero (or low) interest on their card balance for a set period of time, before returning to the standard ``retail'' interest rate.

Setting the retail interest rate of a credit card has been studied extensively, with a risk-based pricing approach being the strategy most commonly used by lenders, and recent research investigating profit-based approaches \citep{Krivorotov_2023}.
However, the optimisation of credit card promotions, which involves tuning many aspects (including the promotion-availability window, the interest-free period duration, and the retail interest rate to charge after the promotion), has received less focus.
\citet{Phillips_2013} describes the optimisation of introductory pricing in order to acquire and retain customers as an open problem, while the recent study of \citet{Kasaian_2022} summarises ``...no study has explored the direct and indirect impact of the teaser rates on bank income.''

Motivated by the lack of focus in the current literature on promotional credit cards, this work aims to explore and understand the behaviour associated with different promotional credit card strategies for lenders and consumers. 
We explore this using an agent-based simulation of promotional and non-promotional credit cards issued by lenders in a simulated environment, modelling the UK credit card market.

This simulation is done using agent-based modelling (ABM), a ``bottom-up'' approach to modelling complex systems (defined by the fine details of a system, rather than large-scale phenomena), with the simulation design consisting of simple ``agents'', their environment, and the interactions that happen between them \citep{Macal_2013}.
Also referred to as agent-based simulations (ABS) or multi-agent systems (MAS), this approach has been used as a tool to understand the evolution of complex systems and to inform decision making across multiple domains, including social, physical, biological, and economic systems \citep{Zhang_2021}.

Experimenting with different pricing techniques in a real-world setting is often costly and impractical, making simulation an ideal tool to explore different strategies.
ABM is a tool that allows us to simulate and model a complex pricing ecosystem without interacting with customers.
More broadly, in finance, agent-based models have been applied to housing markets \citep{Carro_2022}, financial markets \citep{Fischer_2014}, and the interactions between banks, businesses, and customers \citep{Assenza_2015}.
ABM has been used to an extent in exploring pricing in several domains, with the recent review paper of \citet{Kell_2022} surveying many studies exploring bidding strategies and optimal pricing techniques.
The work of \citet{Kell_2022} in particular notes that ABM is a popular technique for pricing applications due to some systems ``acting in unpredictable ways that cannot be predicted a-priori''.
However, the literature using ABM to explore pricing in the banking domain appears to not be as rich as others.
Studies similar to the present work which have used ABM approaches for credit card pricing have been rather sparse, with the works of \citet{Li_2014} and \citet{Nejad_2016} the most relevant.
Only \citet{Nejad_2016} discussed the impact on revenue for the credit card lender, and neither of these studies showed rigorous calibration and validation that would confirm the fidelity of the model to real-world behaviour.

In this work, we introduce a novel agent-based model of a credit card market, which allows for the exploration of the effects of various credit card promotion strategies. 
Our model, distinct from previous agent-based models, places a strong emphasis on replicating realistic behaviour, achieved by using historical distributions of agent attributes and calibrating the model to historical values. 
This approach enhances the fidelity of the model to real-world behaviour, an aspect that has been overlooked in previous agent-based approaches to pricing credit cards.
Furthermore, we provide a comprehensive methodology that applies the model to investigate the impact of different aspects of credit card price promotions, for both the customer and the lender. 
This includes the effects of the interest-free duration offered to customers, the length of the campaign run by a lender, and the impact of competing promotions. 
Our approach provides valuable insights into the dynamics of credit card promotions and serves as a robust framework for further exploration and understanding of lender strategies in the context of credit card promotions.

The subsequent sections are organised as follows.
A review of previous literature is provided in Section \ref{sec:related_work_label}, followed by a description of the model design in Section \ref{sec:methodology_label}.
The calibration of the model to historical values alongside other parameters used in the model is described in Section \ref{sec:calibration_label} and the validation undertaken is described in Section \ref{sec:validation_label}.
Next, we present experiments analysing the effects of different aspects of credit card promotions, the results of which are presented in Section \ref{sec:experiments_label}, followed by our summary and conclusion in Section \ref{sec:summary_label}.
Details outside the main body of the text are provided in Appendix \ref{sec:appendix}.
\section{Related work}
\label{sec:related_work_label}

The pricing of loan products, including credit cards, has been a subject of extensive research, with risk-based and profit-based pricing methods being the most prevalent \citep{Phillips_2019}.
Risk-based pricing, where the retail interest rate offered to a consumer is determined by their perceived risk to the lender, is a common approach for credit cards \citep{Furletti_2003}.
However, recent studies have highlighted that the risk and profitability of customers do not have a monotonic relationship, leading to a shift towards profit-based pricing \citep{Krivorotov_2023}.

For a revolving credit product, like a credit card, the main source of revenue comes from charging interest on revolving card balances (although in some markets, annual fees contribute notably to lender revenue) \citep{Adams_2022}.
Due to this, the consequences for a lender will strongly depend on the behaviour of the borrower post-acquisition, including the extent to which their line of credit is used and how quickly it is paid off, adding a significant amount of complexity to the outcome of a credit card whenever it is taken out \citep{Phillips_2013}.

In addition to the initial price quoted at acquisition, lenders may use pricing techniques to determine subsequent changes to the interest rate or available line of credit \citep{Trench_2003}.
This is a common approach in credit card and other revolving products to increase customer retention in competitive markets \citep{Phillips_2019}, increasing lender revenue.

As noted in \citet{Phillips_2013}, there has been much work on acquisition pricing which involves determining a price to offer to a customer applying for a loan or credit product.
There has also been some work on retention pricing \citep{Trench_2003, Meko_2011}.
However, despite their ubiquity in credit card markets and their ability to increase long-term lender revenue, the optimisation of introductory promotions 
has been notably neglected by the literature.
Lenders use these low introductory or ``teaser'' rates to attract new customers and 
only charge interest on the remaining credit balance whenever the initial promotion has expired.
The recent study of \citet{Kasaian_2022} notes the lack of focus on how teaser rates impact lender income and \citet{Phillips_2013} describes the optimisation of introductory pricing in order to acquire and retain customers as an open problem.

Price promotions are notable as the behaviour of competing lenders plays a large role in the usage of a credit product, with shopping around for the best available deal easier than ever for customers, thanks to easily accessible resources like price comparison websites \citep{FCA_2015}.
As the offers from credit card providers are available to consumers before an application is made, competition cannot be neglected as a factor when aiming to effectively optimise credit card acquisition and retention strategies.
However, competition is often not included in the pricing optimisation systems employed in financial services, as noted by \citet{Phillips_2013}.

Other possible approaches to modelling optimal strategies in the presence of competitors include game theory, which models the interaction between rational agents, with the reward and optimal strategy for one player dependent on the actions of other players in the game.
This approach has been used previously in some pricing optimisation research \citep{Sheha_2020} but has not been widely adopted for credit card pricing strategies due to the complex and often non-rational behaviour of credit card customers \citep{Kasaian_2022}.

Machine learning has received notable attention in recent years as a new way to correctly assess customer risk,
with \citet{Krivorotov_2023} finding machine learning models effective in targeting the most profitable customers.
The use of reinforcement learning of agents in simulated environments is well known through video game challenges \citep{Arulkumaran_2017}.
To the best of our knowledge, the application of reinforcement learning in determining pricing strategies for consumer credit products has been limited, with the exception of \citet{Khraishi_2022}, who applied a model-free offline deep $Q$-learning algorithm for pricing consumer credit, and \citet{Trench_2003}, who developed a model-based reinforcement learning approach for credit card pricing.

ABM offers a promising approach for studying complex systems where traditional "top-down" methods fall short.
As well as replicating the emergent phenomena where other approaches fail, ABM allows understanding of which interactions and processes contribute to emergent effects.
ABM can be effectively applied for the exploration of hypothetical ``what-if'' scenarios (e.g., the consequences of policy changes) and the generation of synthetic data where sourcing of such data may not be feasible.
Across several domains, ABM has been mainly used as a framework to understand the behaviours leading to emergent macroscopic behaviours \citep{Martinez-Gil_2017} to aid and inform the decision making processes of individuals \citep{Balke_2014} and organisations \citep{Yun_2015}.

Several areas of finance and banking have received notable attention from an ABM perspective.
An article by \citet{Farmer_2009} shortly after the economic crash of 2008 criticised the ``top-down'' approaches many institutions were using to assess stability of large economies and suggested ABM was a crucial tool to correctly assess risks to market stability.
Since then, many agent-based models have been developed to understand the causes and effects of instability in the wider banking system \citep{
Liu_2020}.
Agent-based models have also been used to study housing markets, with a particular focus on the causes of growth of housing bubbles, as was the case in the ABM studies of \citet{Axtell_2014} and \citet{Ge_2017}.
An agent-based model of the UK housing market developed by the Bank of England allowed investigations of the effects of regulating mortgage loan-to-value (LTV) and loan-to-income (LTI) ratios on the magnitude of house price cycles \citep{
Carro_2022}.
Using agent-based models for financial market simulations has also received attention as a method of investigating different trading strategies \citep{
Eloubani_2021}.

Whilst literature of applications of ABM to pricing is generally sparse, there are some notable exceptions in various industries and product markets.
In an early work, \citet{North_2009} developed an ABM framework for consumer packaged goods markets in order to support marketing decisioning.
Following this, \citet{He_2014} also developed an agent-based model of a manufacturing distributive system, with a specific focus on determining optimal pricing and location strategies in the presence of price-elastic and price-inelastic products.
In \citet{Du_2019}, an agent-based model of a distributive system is once again considered to determine optimal pricing strategies, exploring differentiated and uniform pricing strategies.
Pricing policies for local and national energy markets have been explored through an agent-based approach in many studies, including \citet{
Rashedi_2016}.
These papers primarily focus on bidding strategies, with price and demand forecasting, and risk and micro-grid management also studied \citep{Kell_2022}.

The simulation of personal credit card usage has been relatively limited.
For example, in one of the few works to investigate this, \citet{Li_2014} use an agent-based approach to model a credit risk system, recommending that lenders carefully evaluate the risk associated with customers who will consider going into an overdraft.
However, this work does not calibrate its model to any historical sources or validate model behaviour against external data.
In addition, as noted by the authors, the work does not take into account the impact on revenue for the lender.
In their studies of the use of debit and credit cards, \citet{Alexandrova-Kabadjova_2008} and \citet{Alexandrova-Kabadjova_2013} use an agent-based network to simulate the adoption and usage of debit and credit cards through the interactions of customers and merchants.
However, these two works focus on potential profitability for merchants charging interchange fees, rather than interest charged to credit card accounts (that make up the bulk of card lender revenue) and do not consider the effect of promotional offers.

The most similar work to this paper is that of \citet{Nejad_2016}, who use an ABM approach to explore the effect of different promotion characteristics (promotional rate and interest-free duration) on lender profit.
However, this model 
assumes that the main motivation for individuals taking out promotional credit cards is through word-of-mouth.
This neglects customer needs for convenient credit or desire to build up a credit score, which are more popular motivations for acquiring a credit card\footnote{
\href{https://www.money.co.uk/credit-cards/why-do-people-use-credit-cardswhy-do-people-use-credit-cards}{https://www.money.co.uk}}, as well as the effect of price comparison websites and bank marketing. 
Second, no calibration to historical values is undertaken by the model, meaning agent behaviour may not be realistic.
Third, there is no validation of the model results, meaning no conclusions can be drawn about the fidelity of the model to real life.
Our work also takes an agent-based approach to exploring credit card promotions, but differs in several ways.
Rather than having the spread of knowledge by word-of-mouth as the main motivation for customers taking out new credit cards, the agents in our model are motivated by a need for credit and a superior offer to the card(s) they currently have.
In addition, our model is calibrated to historical benchmarks, indicating it can reproduce some real-life behaviour.
Finally, our model is validated in a multi-faceted way, by comparison to stylised facts and time-wise data, meaning its fidelity to real-life behaviour can be assessed much more deeply.
\FloatBarrier
\section{Agent-based model description}\label{sec:methodology_label}
\subsection{Model overview}

Our agent-based model comprises two types of agents: customers and lenders. 
These agents interact primarily through credit card applications, card usage, and balance repayments, which we detail in the following subsections.
We base our model on the UK credit market due to the availability of data from reliable sources, such as government bodies and credit bureaus. 
However, our model can be adapted to represent any credit card market.

Each customer agent $j$ can possess a maximum of $\max_cards$ cards, denoted as $\customer_cards_j$, where $0 \le \customer_cards_j \le \max_cards$.
Each card $k$ held by customer $j$ has fixed attributes of credit limit $\credlimit_{j, k}$ and interest rate $i_{j, k}$, which are set by the lender of the card.
The card balance $0 \le \balance _{j, k} \le \credlimit_{j, k}$ is a dynamic attribute.
For simplicity of notation, we drop the time index for this variable (and all other time-dependent variables).
As a simplifying assumption, credit card annual fees are not included in our current model, although they are an important component of income for some lenders. 
We use a monthly time step $t$ as credit card billing and repayment schedules occur on a monthly level and run for $T > 0$ time steps, where $T$ is the maximum number of time steps the model will run for.

The model is implemented in Python, using the Mesa framework \citep{mesa_2020}.
Customer and lender agents are defined by their attributes, behavioural parameters, and possible actions, which we describe in more detail in the following subsections.
The data sources used to determine credit card behaviours are described in Section \ref{subsec:data_sources_label}, for both estimated parameters (those that can be directly taken from sources) and calibrated parameters (those that are determined through calibration).
All of the agent attributes and behavioural parameters are listed in Appendix \ref{subsec:customer_attributes_label}.

\subsection{Customer actions}\label{subsec:customer_actions}
At each time step $0 \le t < T$, all customer agents can choose from several possible actions, or opt to take no action at all.
If customer $j$ has $\customer_cards_j = 0$, their only possible action for the turn is to apply for a credit card. 
If a customer has $0 < \customer_cards_j < \max_cards$, their possible actions for the turn are: repay credit card(s), apply for another credit card, or use credit card(s). 
If a customer has $\customer_cards_j = \max_cards$, their possible actions for the turn are: repay credit cards or use credit cards.
These possible actions are further explained in the following sections.

\paragraph{Customer action 1: Apply for a credit card}
This action is available to customers who hold fewer than the maximum number of credit cards and consists of two steps.
First, the customer decides whether to apply for a credit card, followed by an application to a chosen lender (if they have decided to apply).

One of the primary motivations for customers to apply for a credit card is to spread the payment of large purchases.\footnote{
\href{https://www.money.co.uk/credit-cards/why-do-people-use-credit-cards}{https://www.money.co.uk}}
To account for this, we define the probability of customer $j$ applying for their first credit card to be related to the difference in their expenditure $\total_exp_{j} > 0$ and their income $\income_j > 0$:
\begin{equation}\label{eq:apply_no_card}
	\mathbb{P}(\text{apply}\mid \customer_cards_j=0) = \text{max}(\sigma(\lambda_{\text{tend}} \cdot (X_{j} - \income_j)), \appminprob ),
\end{equation}
where $\sigma$ is the sigmoid function, $\sigma(x) = 1 / (1+e^{-x})$, $\apptend \in \mathbb{R}$ is a calibrated parameter that controls the relationship between a customer’s likelihood to apply and the difference between their total expenditure and income, and $0 \le \appminprob \le 1$ represents the minimum probability at each step that a customer will apply for a credit card.
		
Customers are also motivated to apply for credit cards to manage their balances\footnote{ibid.} and to take advantage of promotional offers \citep{Murthi_2019}.
To account for this, the probability of applying for additional cards is dependent on how close a customer is to reaching their total credit limit and the difference in the expected cost of their current card(s) and potential other cards.
The cost of card $k$ owned by customer $j$ is shown by $\expected_cost_{j, k} \ge 0$ and the expected cost of the card offered by lender $l$ is denoted $\widetilde{\expected_cost}_{l} \ge 0$.
Thus, for customer $j$ with $1 \le \customer_cards_j \le \max_cards - 1$, we define their probability of applying as having a component related to their credit card utilisation, $\utilisation_component_j \ge 0$, and a component related to offers available, $\offer_component_j \in \mathbb{R}$:
\begin{equation}\label{eq:apply_more_one_card}
		\mathbb{P}(\text{apply}\mid 1 \le \customer_cards_j \le \max_cards - 1) = \sigma(\utilisation_component_j + \offer_component_j - \appconst),
\end{equation}	 
where
\begin{equation}\label{eq:apply_more_one_cred_lim}
		\utilisation_component_j = \applimitweighting \cdot \frac{1}{\customer_cards_j} \sum_k \frac{\balance_{j, k}}{\credlimit_{j, k}},
\end{equation}
\begin{equation}\label{eq:apply_more_one_card_offer}
	\offer_component_j = \better_offer_weighting \cdot \left( \frac{1}{\customer_cards_j} \sum_{k} \expected_cost_{j, k} - \text{min}_{l}(\widetilde{\expected_cost}_{l}) \right),
\end{equation}

where $\applimitweighting > 0$ relates how likely a customer is to apply for a new card and how close on average they are to the credit limits of the cards they own, $\better_offer_weighting > 0$ controls the effect of offers compared to a customer's current credit card(s) on the likelihood of them applying to a new one, and $\appconst \is_real_number$ controls the overall probability of customers applying to further credit cards.
Note that $\frac{1}{\customer_cards_j} \sum_{k} \expected_cost_{j, k}$ refers to the mean expected cost from $\customer_cards_j$ cards owned by customer $j$.

Customer agents estimate the expected cost of cards by calculating the interest that would be charged for a representative balance on the card over a fixed horizon, in a similar way to how comparison websites compare cards.
In our simulations described in later sections, we use a representative balance of £1,200 and a fixed horizon of 48 months.
	
If a customer decides to apply for a credit card then they must decide which of the credit cards on offer to apply for. 
Cards are first ranked by their estimated cost to the customer.
We account for the observation of \citet{Agarwal_2015} that customers do not always make the optimal decision when choosing from credit cards available to them, and so define the probability of customer $j$ making the rational application choice as dependent on the difference in the expected cost of the most expensive card they possess and the expected cost of the cheapest card offered by lenders:
\begin{equation}\label{eq:prob_rational_application}
P^{(\text{r})}_{j} = \mathbb{P}(\text{rational choice}) = \sigma(\apprationality \cdot (\text{max}_k(\expected_cost_{j, k}) - \text{min}_l(\widetilde{\expected_cost}_{l}))),
\end{equation}
where $\lambda_{\text{r}} > 0$ is a calibrated parameter that controls the likelihood of a customer making a rational decision.

A customer $j$ has a probability $P^{(\text{r})}_{j}$ of applying to the card on offer with the lowest expected cost, and a probability ($1 - P^{(\text{r})}_{j}$) of applying to a card on offer at random.
Note that when customers are presented with identical offers, they will pick between offers with equal probabilities.
The sigmoid function used in this action is an approximation that may be improved upon in future model iterations by a more complex function.	

\paragraph{Customer action 2: Use credit card(s)} This action is available to any customer agent with $\customer_cards \ge 1$. 
The amount spent across the credit card(s) owned by customer $j$ is determined by the usage tendency multiplied by the available expenditure and how close they are to their overall credit limit:
\begin{equation}\label{eq:spend_calc}
	CRED_j = \usage_tendency \cdot \credexp_{j} / 12 - \cred_lim_weighting \cdot \left(\sum_k \balance_{j, k} / \sum_k \credlimit_{j, k} \right),
\end{equation}
where $\theta_{\text{tend}} > 0$ is a calibrated parameter that controls the effect of creditable expenditure $\credexp_j > 0$ on the customer agent's card use and $\cred_lim_weighting > 0$ relates overall card utilisation to a customer agent's card use.
Creditable expenditure is defined as the component of a customer's expenditure that can be paid for using a credit card.
The categorization of expenditure as either creditable or non-creditable is defined in Appendix \ref{subsec:exp_categories}.
	
If a customer holds a single credit card, then they can only use that card.
Subsequently, the user's credit card balance increases by $CRED_j$ amount.
If a customer holds multiple credit cards then the agent makes a decision on how to spread payment amongst these cards.
Customers determine the usage order based on which cards have the lowest interest rates and balances, as found by \citet{Murthi_2019}.

To model the usage order, we assign weight $w_{j, k} > 0$ to each card $k$ owned by customer $j$.
The weights are determined as follows:
\begin{equation}\label{eq:spend_weighting}
	w_{j, k} = \mu_{\text{i}} \cdot \credlimit_{j, k} + \mu_{\text{\balance}} \cdot \balance_{j, k},
\end{equation}

where $\mu_{\text{i}} > 0$ and $\mu_{\text{\balance}} > 0$ are sensitivity parameters calibrated to weight the importance of interest rates against balances, as well as to account for scaling. 

The weights of each card calculated using Equation (\ref{eq:spend_weighting}) determine the order of credit cards to use.
Customers iterate through these cards, filling each up to its comfort limit, i.e. the proportion of the credit limit a customer is comfortable using.
Customers are generally advised not to use more than 30\% of their credit limit\footnote{
\href{https://experian.co.uk/consumer/credit-cards/guides/credit-limits.html}{https://experian.co.uk
}
}, so we simplify their spending behaviour and assume customers will spread payments until they have to exceed this.
If customers have not used up all of $CRED_j$, they iterate through their cards in order again based on the weights.
This time each card is filled up to the credit limit (which the balance of a card can never exceed), until $CRED_j$ is exhausted. 

\paragraph{Customer action 3: Repay credit card(s)}
This action is available to any customer agent with $\customer_cards \ge 1$.
In our model, we incorporate the correlation between customer credit scores and regular repayments, a relationship that is well-documented in the literature.\footnote{
\href{https://experian.co.uk/consumer/guides/what-affects-score.html}{https://experian.co.uk
}
} Specifically, we model the budget that a customer agent $j$ has to repay their credit card(s) each turn as being influenced by their credit score, $\score_j$:
\begin{equation} \label{eq:payment_calc}
		BUDGET_j = \repaytend \cdot \score_j + \repayconst,
\end{equation}
where $\repaytend > 0$ is a calibrated scaling parameter that controls the relationship between a customer's credit score and their balance repayment and $\repayconst$ controls how much of their credit card balance customers can repay.

If a customer holds a single credit card $k$, their card balance subsequently decreases by $BUDGET_j$ (to a minimum balance of zero):
\begin{align}
\begin{split}
	PAY &= \text{min}(BUDGET_j, \balance_{j, k}),
	\\	
	\balance_{j, k} & \gets \balance_{j, k} - PAY.
\end{split}
\end{align}

For customers holding multiple credit cards, we model repayment behaviour of them based on the study from \citet{Gathergood_2019}, which found 15\% of customers repaying the card with the highest interest rate first (avalanche strategy), 10\% prioritising the lowest rate card (anti-avalanche strategy), and the remaining 75\% of customers randomly making payments across their cards.
We use this distribution to assign a repayment strategy to each customer agent, guiding the priority of their card payments every turn they take a repayment action.

Based on the assumption that customers always make the minimum payment on their card(s) if they can afford it, customers first iterate through their ordered cards, making the minimum payment $M_{j, k} \ge 0$ (determined by the lender of the card on the previous step) on each card, stopping if they have exhausted $BUDGET_j$.
There is a finite probability that a customer misses making the minimum payment on each card on each iteration, $P_{\text{miss}}$, which is a calibrated parameter.
Thus, customer $j$ with cards $k\in\{0, 1, 2,..,\customer_cards_j\}$ rank-ordered based on their assigned repayment strategy undertakes the procedure described in Algorithm \ref{alg:min_payments} in Appendix \ref{subsec:repayment_algo}.

Note that if a customer does not make a payment on a card on this first iteration through their cards, they can still make a payment in the subsequent iteration and avoid a missed payment fee.

The customer will then iterate through their cards in the order they have determined, paying off as much of the balance on each card as they can (to a minimum balance of zero), until $BUDGET_j$ is exhausted.
Pseudocode for this procedure is show in Algorithm \ref{alg:repayment} in Appendix \ref{subsec:repayment_algo}.

To approximate the process undertaken whenever borrowers fall into arrears, customers who do not make at least the minimum payment on a credit card three months (time steps) in a row will ``default'' on their credit card.
When this happens, the customer agent cannot use or repay the card or re-apply for the card.
The lender of the card is not able to charge interest or fees on the card and the balance on the card is deducted from the lender profits.

\subsection{Lender actions}\label{sec:lender_actions} 
At each time step, lenders will undertake billing cycle actions, including charging interest and fees, for customers who have taken out one of their credit cards.
If a lender has given out no credit cards, or all of their customers have defaulted on their cards, the lender takes no actions.
Credit cards offered by lenders are defined in the initial conditions of the model, along with the time steps at which credit card promotions run.

\subsubsection{Lender credit card underwriting}
We assume that at each time step, lenders may accept or reject credit card applications as customers make them.
As a simplification of the underwriting process, upon receiving an application from a customer agent, the lender decides whether to accept or decline each application based on the applicant's income and credit score being above set thresholds.
If successful, the customer is provided with a credit card, for which a credit limit must be determined for the card.
The criteria for credit limits given by lenders are not widely accessible, but it is known that income is a strong factor in the credit limit given to a customer.\footnote{
\href{https://experian.co.uk/consumer/credit-cards/guides/credit-limits.html}{https://experian.co.uk}
}
Following this, we approximate the credit limit provided to customer $j$ for card $k$ as a logarithmic function $\credlimit_{j, k} = \mathrm{A} \cdot \log(\income_j/\mathrm{B})$, where $\mathrm{A}$ acts as a linear scalar of the limit offered and $\mathrm{B}$ controls the growth rate for the limit as a function of customer income offered.

\subsubsection{Lender billing cycle actions}
In our model, following the completion of all customer agent actions for each step, each lender takes the following actions in order:
\paragraph{Process balance statements} The lender creates a balance statement for each credit card associated with it, based on the amount on the card's balance (see Section \ref{subsec:estimated_parameters}).
The lender also determines the minimum payment for a customer as a fixed fraction of their card balance.
If this amount is not paid the next month, a fee is charged to the card.
\paragraph{Charge interest and fees}
The lender charges interest on any outstanding balance across a customer's credit card held with the lender. 
Outstanding balance is any balance that has not been repaid according to the credit card's billing schedule.
If the most recent payment on the card is less than the minimum payment, the lender charges a fixed fee to the customer's account.

\subsection{Implementation of the model}
Each simulation is initialised with a user-defined number of customer and lender agents, set to run for $T$ time steps.
Customer agents have properties sampled from historical distributions (as described in Section \ref{subsec:data_sources_label}), with estimated and calibrated parameters pre-defined for the simulation (see sections \ref{subsec:estimated_parameters} and \ref{subsec:calibrated_parameters} for more details).
At each time step $t$, all customer agents undertake their actions, followed by all lenders taking their actions.
The simulation continues until $t$=$T$, with the properties of customer and lender agents recorded each time step.

\section{Estimation and calibration of model parameters}\label{sec:calibration_label}
In line with the categorisation proposed by \citet{Carro_2022}, we classify the parameters that govern the attributes and actions of agents (as detailed in Appendix \ref{subsec:customer_attributes_label}) as either estimated or calibrated parameters.
Estimated parameters are values and distributions that can be directly obtained from available data sources.
In contrast, calibrated parameters are those that cannot be directly derived from available sources and are typically associated with the psychological decision-making processes of agents.
These parameters are determined through model calibration to historical values, using the method of simulated moments \citep{Franke_2009}.

\subsection{Data sources}\label{subsec:data_sources_label}
Data was obtained from historical sources to inform agent property distributions, for estimated parameters, for historical values used to calibrate the model, and for model validation.
Where possible, data was sourced from late 2019 to early 2020, motivated by the model replicating a modern credit card market, while avoiding the disruptive effects of the COVID-19 pandemic and lockdowns.
Sources based on populations in the UK were prioritised, as were established institutions (e.g. government bodies), and sources which evidenced their methodology.
Further details of these sources and how they were used can be found in Appendix \ref{subsec:customer_attributes_label}.

We derived the distribution of customer ages from a 2020 survey by the Office of National Statistics (ONS) of the UK population \citep{ONS_Age_2020}.
A survey of UK gross income across age groups from the ONS allowed us to assign a gross income to customer agents.
We used UK government websites to source income tax \citep{Income_tax_2022} and national insurance contributions \citep{National_insurance_2022}, enabling us to convert gross income into net income.
The ONS Living Costs and Food Survey \citep{ONS_Expenditure_2020} provided average expenditure per income decile for the UK population, broken down by expenditure categories, allowing us to calculate total and creditable expenditure for individuals.
We could not find public data on UK credit card scores, so we used data from Statista in 2019 for 267 million U.S. Americans \citep{Statista_credit_score} to assign credit scores to customer agents as a function of age.

As detailed below, we sourced estimated parameters and historical values used for calibration for the model from relevant UK sources from late 2019 or early 2020.
For the time-wise validation of the model described in Section \ref{sec:validation_label}, we used data from Virgin Money UK annual reports, starting from the launch of their credit card business in 2012.

\subsection{Estimated parameters}\label{subsec:estimated_parameters}
Table \ref{tab:estimated_parameters_table} summarises the estimated parameters in the model.
We determined customer agent attributes using publicly available datasets, as described in Section \ref{subsec:data_sources_label}.
These attributes include customer ages, incomes, expenditure amounts, and credit scores, with histograms provided in Appendix \ref{sec:appendix}.
Unless specified otherwise, the retail interest rates of cards, $i$, is set to 20\% to approximate typical pre-COVID UK credit card interest rates.\footnote{\href{https://www.money.co.uk/credit-cards/why-do-people-use-credit-cards}{https://www.money.co.uk}}
We derived the distribution of customers following certain repayment strategies from studies of historical customer behaviour \citep{Gathergood_2019}.
Where possible, these values were taken from late 2019 or early 2020, to avoid the disruptive effects related to the COVID-19 pandemic.
For all lenders, the acceptance criteria is customer annual income exceeding £7,500 and a credit score in excess of 600.
While there is no sharp threshold at which lenders will accept a card application, incomes from £7,000 to £7,500 are considered the lowest income at which a lender might offer a card at\footnote{\href{https://moneysupermarket.com/credit-cards/can-i-get-a-credit-card/}{https://moneysupermarket.com}}
 and FICO scores below 600 are considered ``bad'' and therefore unlikely to be accepted by lenders.\footnote{\href{https://www.sofi.com/learn/content/minimum-credit-score-for-credit-card/}{https://www.sofi.com}}
The parameters $\mathrm{A}$ and $\mathrm{B}$ that control the credit limit offered to a customer are chosen to give a mean credit limit consistent with the average limit in the UK which is around £3,000 to £4,000.\footnote{
\href{https://www.finder.com/uk/credit-card-limit}{https://www.finder.com
}
}
For simplicity, we set the maximum number of cards that a customer can own in the model to be $N$=3 (although simulations with larger numbers of credit cards are possible).

\begin{table*}[ht]
\centering
\begin{threeparttable}
\caption{Estimated model parameters.}
\label{tab:estimated_parameters_table}
\centering
\begin{tabular}{lrl}
\hline
Parameter  &   Value & Source \\ \hline
Retail APR & 20.0\% &  \tnote{a} \\
Late payment fees &  £12.00 & \tnote{b} \\
Avalanche repayment strategy fraction & 0.15 & \citet{Gathergood_2019} \\
Anti-avalanche repayment strategy fraction & 0.10 &  \citet{Gathergood_2019}\\
Random repayment strategy fraction & 0.75 &  \citet{Gathergood_2019}  \\
Percentage of balance for minimum payment & 2.5\% & \tnote{c} \\ 
Linear scalar of credit limit offered ($\text{A}$) & \pounds 6000 & \tnote{d} \\ 
Growth rate of credit limit offered ($\text{B}$) & \pounds 6000 & \tnote{d} \\ 
\hline
\end{tabular}
\begin{tablenotes}[0.8]\footnotesize
\item [a] \href{https://www.nimblefins.co.uk/average-credit-card-interest-rate-apr-uk}{https://www.nimblefins.co.uk}
\item [b] \href{https://www.money.co.uk/credit-cards/why-do-people-use-credit-cards}{https://www.money.co.uk}
\item [c] \href{https://www.halifax.co.uk/creditcards/help-guidance/minimum-payments.html}{https://www.halifax.co.uk}
\item [d] \href{https://www.finder.com/uk/credit-card-limit}{https://www.finder.com}
\end{tablenotes}
\end{threeparttable}
\end{table*}

We approximate the number of monthly new entrants to the simulation to be 0.091\% of the customer agent population, taken as a 12\textsuperscript{th} of the population of the UK aged 17 in 2020.\footnote{
\href{https://www.ons.gov.uk/peoplepopulationandcommunity/populationandmigration/populationestimates/datasets/populationestimatesforukenglandandwalesscotlandandnorthernireland}{https://www.ons.gov.uk}
}
To maintain a steady agent population, an equivalent number of customers are randomly selected for churn from the model at each time step.

\subsection{Calibrated parameters}\label{subsec:calibrated_parameters}
Calibrated parameters were determined using the method of simulated moments, as performed by \citet{Franke_2009}, \citet{Chen_2018}, and \citet{Carro_2022}.
This involves constructing an objective function based on the differences between a set of moments of historical attributes and their corresponding values from a model simulation.
Running simulations across the possible parameter space to find the minimum value of the objective function indicates the calibrated parameters to be used for future experimental runs.
The mean absolute percentage error (MAPE) between the true and model values of simulated moments was chosen as the objective function due to its scale-independence and sensitivity to outliers.

\begin{table*}[ht]
\caption{Calibrated model parameters.}
\label{tab:calibrated_parameters_table}
\centering
\begin{tabular}{llr}
\hline
Parameter & Equation & Value \\ \hline

Application constant ($\rho$)  & Eqn. \ref{eq:apply_no_card} & 5.0   \\
Application tendency ($\apptend$) & Eqn. \ref{eq:apply_no_card} & 0.001\\
Application minimum probability ($\appminprob$) & Eqn. \ref{eq:apply_no_card} & 0.08 \\
Application rationality ($\apprationality$)  & Eqn. \ref{eq:prob_rational_application} & 0.007   \\
Application limit weighting ($\applimitweighting$)  & Eqn. \ref{eq:apply_more_one_cred_lim} & 3.0 \\
Better offer weighting ($\better_offer_weighting$)  & Eqn. \ref{eq:apply_more_one_card_offer} &  $4.0\times10^{-5}$ \\
Usage tendency ($\usage_tendency$) & Eqn. \ref{eq:spend_calc} & 1.8  \\
Repayment tendency ($\repaytend$) & Eqn. \ref{eq:payment_calc} & 0.024  \\
Repayment constant ($\repayconst$) & Eqn. \ref{eq:payment_calc} & 1,700.0   \\
Interest rate payment weighting ($\interestweighting$) & Eqn. \ref{eq:spend_weighting} & 0.5 \\
Balance payment weighting ($\balanceweighting$) & Eqn. \ref{eq:spend_weighting} & 0.9 \\
Credit limit weighting ($\cred_lim_weighting$) & Eqn. \ref{eq:spend_calc} & 2,000.0 \\
Missing min. payment probability ($\miss_pay_prob$) & - & 0.0012 \\ \hline

\end{tabular}
\end{table*}

Values of the moments were taken from late 2019 to early 2020 where possible, so as to use data that was recent but before COVID-19 lockdowns, which dramatically affected consumer use of credit cards.
Additionally, this time frame is consistent with the period of the data sources used to create the distributions of agent attributes.

Calibration took place across models containing 5,000 customer and 3 lender agents, with a single lender undertaking a promotion and $T$=120.
Manual exploration of the calibration parameter space found approximate ranges for the calibrated parameters, followed by exploration using the Optuna framework and a Tree-Structured Parzen Estimator \citep{optuna_2019}
to efficiently survey the parameter space and minimise the objective function.
The final calibrated values can be found in Table \ref{tab:calibrated_parameters_table}.
Using a simulation of 30,000 agents across 5 seeds to estimate the calibration performance gave a MAPE of 15.1\:$\pm$\:0.6\%, comparable to that achieved in \citet{Carro_2022}.
The values of historical benchmarks achieved can be found in Table \ref{tab:moments_table}.

\begin{table*}[ht]
\centering
\begin{threeparttable}
\caption{Moments from the calibrated model and their target values. Simulated values are given as the mean across 5 initial seeds, with the uncertainty reported as one standard deviation.}
\label{tab:moments_table}
\centering
\begin{tabular}{lrrl}
\hline
Moment 										& Simulation value & Target (actual) value & Source  \\ \hline
Average monthly account spend  			& £599.10 $\pm$ 1.4			& £655 					& \tnote{a} \\ 
Average account balance 					& £1879 $\pm$ 20		& £1720 				& \tnote{b} \\ 
Overall payments to balance ratio 			& 0.318 $\pm$ 0.004		& 0.32 					& \tnote{b} \\ 
Proportion adults with one or more cards   & 0.7147 $\pm$ 0.0013		& 0.62 					& see footnote \tnote{c} \\
Average no. cards 							& 1.4722 $\pm$ 0.0027		& 1.14 					& \tnote{d} \\ 
Proportion paying interest 					& 0.3163 $\pm$ 0.0024		& 0.36 					& \tnote{e} \\ 
Proportion accounts with missed payments 	& 0.0209 $\pm$ 0.0006		& 0.015 				& \tnote{f} \\ 
Proportion of rational applications  		& 0.5949 $\pm$ 0.0004		& 0.60 					& \cite{Agarwal_2015}  \\ \hline
\end{tabular}
\begin{tablenotes}[0.8]\footnotesize
\item [a] \href{https://www.fico.com/blogs/uk-card-spend-hits-over-two-year-high}{https://www.fico.com}
\item [b] \href{https://www.fico.com/en/newsroom/fico-uk-credit-market-report-november-2020-new-data-raises-concerns-about-post-christmas}{https://www.fico.com}
\item [c] 60.1m cards \cite{Statista_credit_card_no_2020}, 52.98m aged 18 and over in UK from \cite{ONS_Age_2020}.
\item [d] \href{https://www.money.co.uk/credit-cards/credit-card-statistics}{https://www.money.co.uk}
\item [e] \href{https://www.fico.com/en/newsroom/fico-uk-credit-card-market-report-may-2022}{https://www.fico.com}
\item [f] \href{https://www.inews.co.uk/inews-lifestyle/money/saving-and-banking/nearly-a-third-of-credit-card-holders-are-paying-almost-1500-on-annual-interest-alone-1557078}{https://www.inews.co.uk}
\end{tablenotes}
\end{threeparttable}
\end{table*}

\FloatBarrier
\section{Validation}\label{sec:validation_label}

We utilize two distinct validation techniques.
The first technique involves comparing the trends generated by our model with stylised facts from previous studies on credit card data sets.
This technique of comparing model outputs with stylised facts has been used as a validation technique for agent-based models in past research, as seen in the works of \citet{Ashraf_2017} and \citet{Popoyan_2017}.
The second technique applies a time-wise validation of our model against historical data from credit cards offered by Virgin Money UK.
This approach is similar to the one used by \citet{Holm_2018} and \citet{Carro_2022}, who replicated historical price changes in different types of wood in wood markets.
By using these two validation methods, we aim to thoroughly assess both the model's ability to accurately reproduce realistic credit card customer behaviours and its potential as a tool for exploring effective promotion strategies.

\subsection{Stylised facts validation}
The model utilised for calibration, which comprises three lenders (with one applying a promotion) and 30,000 customer agents over a period of $T$=120, is also employed for validation against stylised facts derived from recent research.
The calibrated parameters, as detailed in Section \ref{sec:calibration_label}, are used in the validation simulation.
We examine the properties of the model from the final step of the simulation to determine their alignment with the stylised facts from previous studies of historical credit card data.

\begin{figure}
     \centering
     \begin{subfigure}[t]{0.45\textwidth}
		\centering
		\includegraphics[width=1.0\textwidth]{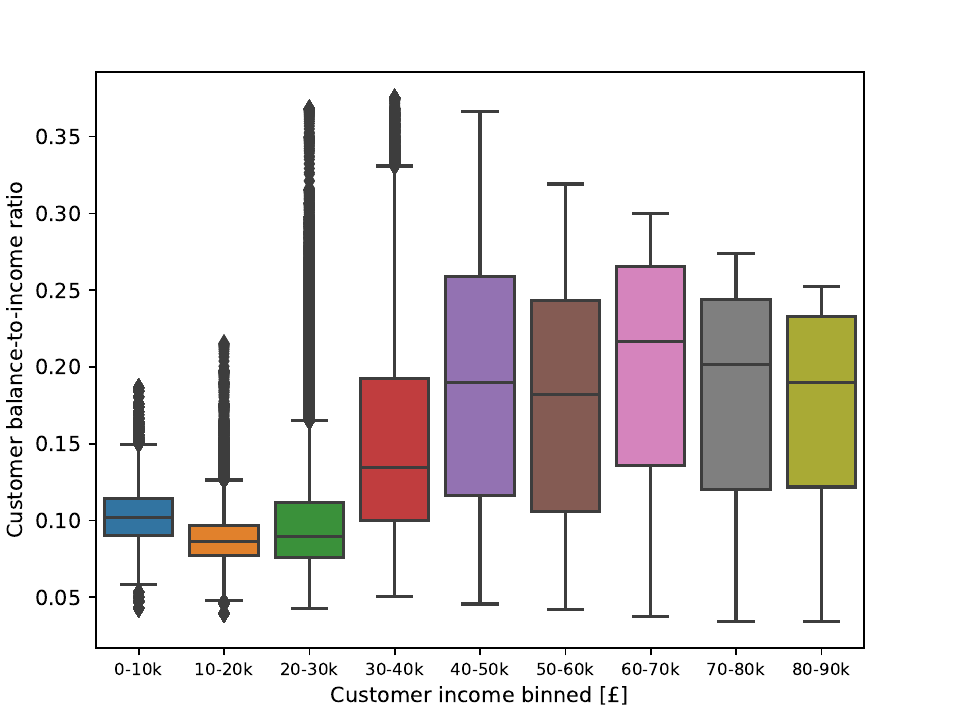}
		\caption{Total card balance-to-income ratio against income for customer agents.}
		\label{fig:Customer_income_Customer_balance-to-income}
     \end{subfigure}
     \hfill
     \begin{subfigure}[t]{0.45\textwidth}
        \centering
        \includegraphics[width=1.0\textwidth]{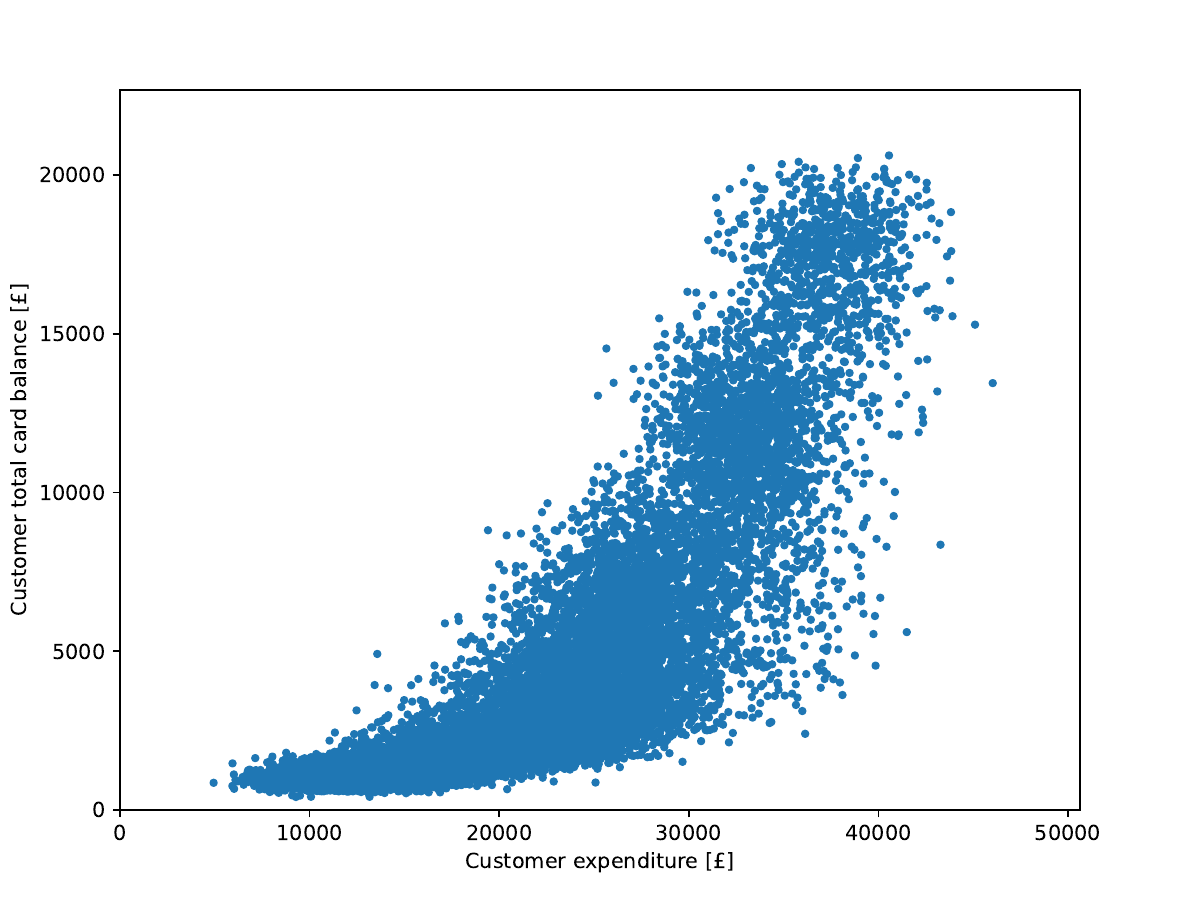}
		\caption{Total card balance against expenditure for customer agents.}
		\label{fig:Customer_expenditure_Customer_total_card_balance}
     \end{subfigure}
     \par\medskip
     \begin{subfigure}[t]{0.45\textwidth}
        \centering
        \includegraphics[width=1.0\textwidth]{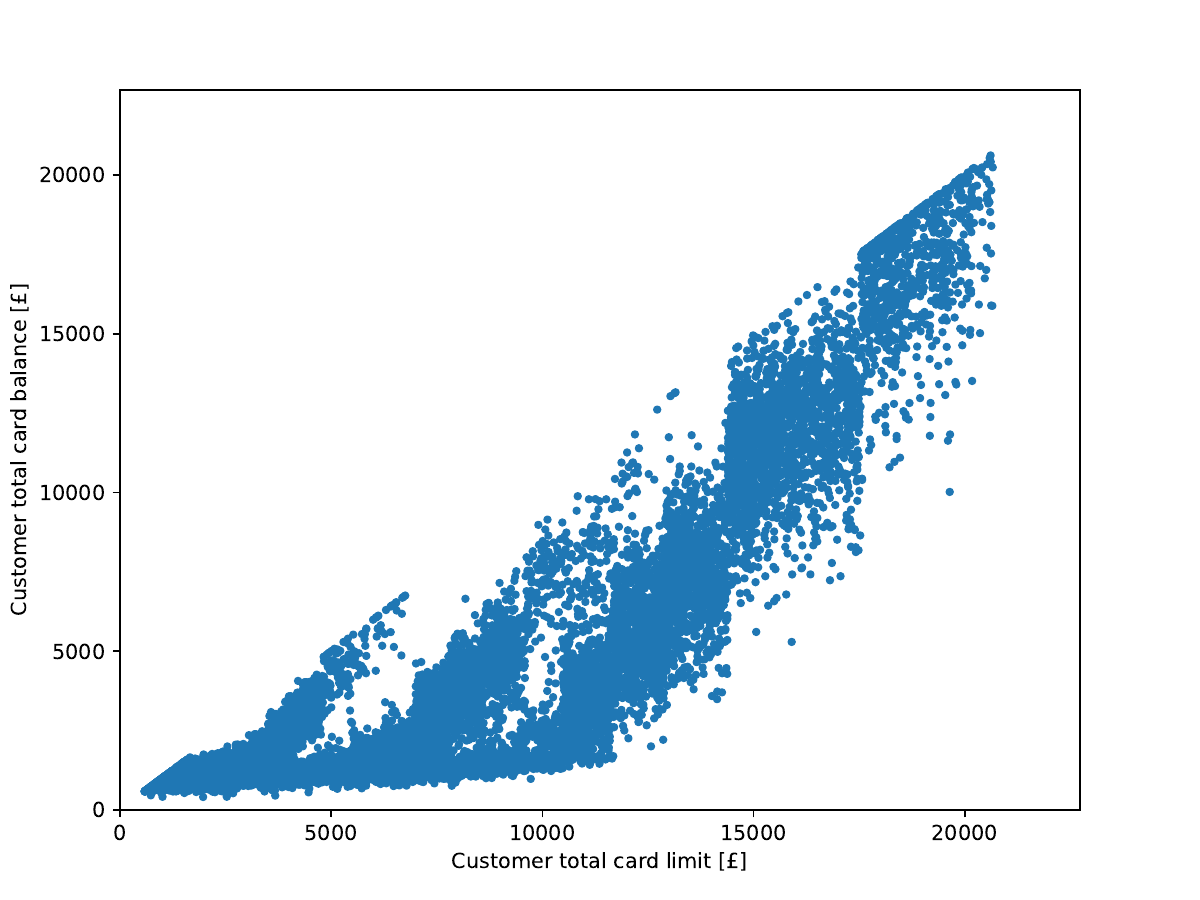}
		\caption{Total card balance against total card limit for customer agents.}
		\label{fig:Customer_total_card_limit_Customer_total_card_balance}
     \end{subfigure}
     \hfill
     \begin{subfigure}[t]{0.45\textwidth}
        \centering
        \includegraphics[width=1.0\textwidth]{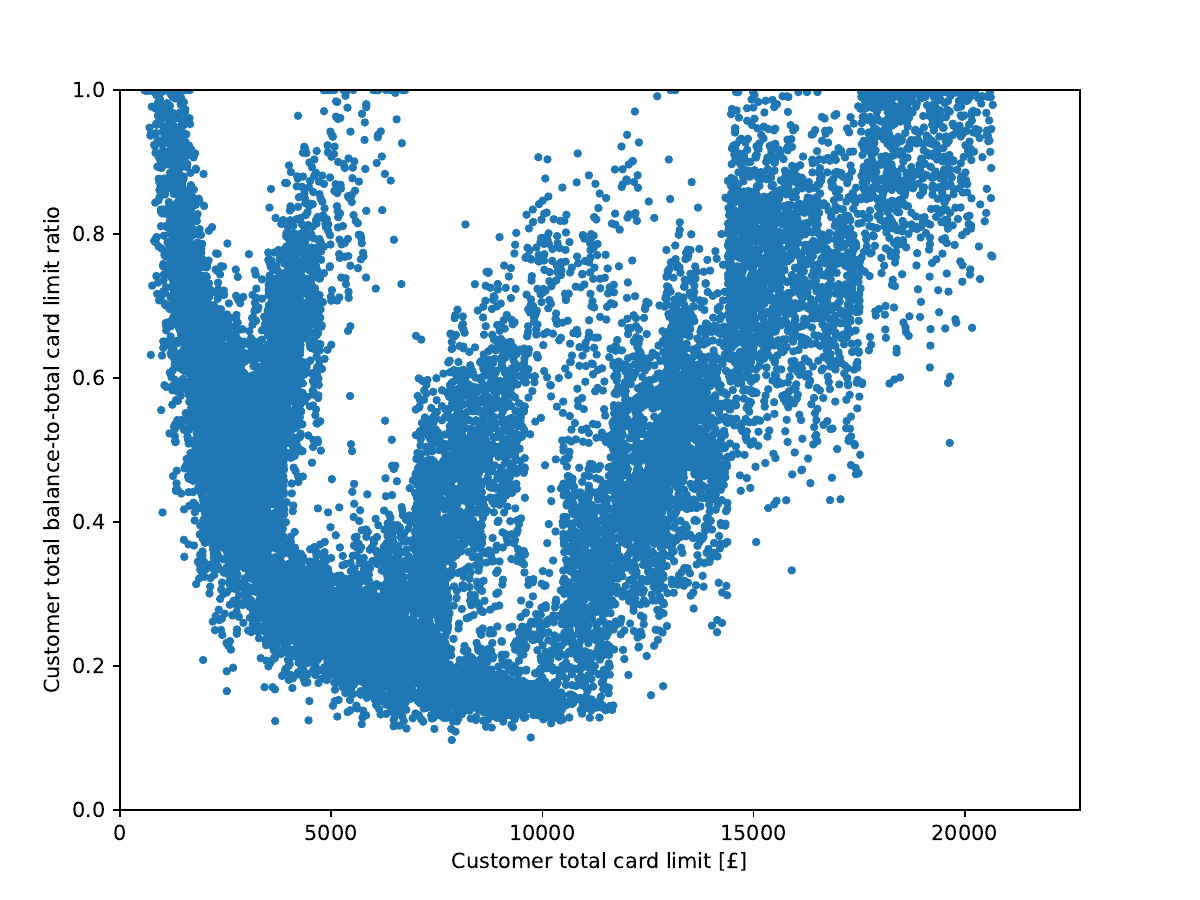}
		\caption{Ratio of total balance to total card limit plotted against total card limit for customer agents.}	
		\label{fig:Customer_total_card_limit_Customer_total_balance_total_card_limit_ratio}
     \end{subfigure}
	 \caption{Plots of relationships between customer agent variables at the final step of a simulation run.
     Attributes shown correspond to stylised facts observed in previous studies.}
     \label{fig:time_wise_validation}       
\end{figure}

We first validate the non-monotonic relationship between customer income and the debt-to-income ratio, as identified by \citet{Kincses_2010}. 
This relationship is particularly pronounced in middle-income households, which display the highest debt-to-income values. 
As illustrated in Figure \ref{fig:Customer_income_Customer_balance-to-income}, this non-monotonic relationship is clearly evident, with middle-income households demonstrating the highest balance-to-income values. 
\citet{Kincses_2010} also found that consumption, or customer expenditure, is a significant driver of credit card balance size. 
This finding is supported by the plot in Figure \ref{fig:Customer_expenditure_Customer_total_card_balance}, which shows a strong correlation between credit card balance and expenditure.

We corroborate our model against another stylised fact, namely the strong positive correlation between balance and credit limits on customers' credit card accounts. 
This correlation was previously observed by \citet{Kasaian_2022}, who conducted an investigation into the effects of credit card ``teaser'' rates on customers from a U.S. bank. 
As depicted in Figure \ref{fig:Customer_total_card_limit_Customer_total_card_balance}, our model also demonstrates this positive correlation, with the total card balance of a customer consistently increasing with the total credit limit. 

Furthermore, in their study of a credit bureau data set, \citet{Fulford_2018} observed that customer credit utilisation, defined as the balance-to-credit limit ratio of a customer, is independent of the customer's credit limit. 
Our model aligns with this finding as well, as shown in Figure \ref{fig:Customer_total_card_limit_Customer_total_balance_total_card_limit_ratio}. 
Here, the ratio of total card balance to total card limit is plotted against the total card limit for customer agents, revealing no discernible strong linear relationship. 

The alignment of our model's output with these stylised facts suggests its fidelity to the typical behaviour of credit card customers.

\FloatBarrier

\subsection{Time-wise validation}
The model's time-wise validation is conducted to evaluate its ability to accurately replicate the outcomes following the introduction of a new promotional credit card product.
The historical data utilised for this validation is sourced from the annual reports of Virgin Money UK\footnote{
\href{https://www.virginmoneyukplc.com/investor-relations/results-and-reporting/annual-reports}{https://www.virginmoneyukplc.com
}
}, selected due to its instance of launching a promotional card during a relatively stable period in the UK market, and the public availability of data for this period.
Following the acquisition of MBNA\footnote{\href{https://www.reuters.com/article/virgin-mbna-idCNL6N0ANCNX20130118}{https://www.reuters.com
}
}, Virgin Money UK established a credit card division and subsequently launched a promotional credit card three years later.\footnote{
\href{https://www.thisismoney.co.uk/money/cardsloans/article-3205864/Virgin-Money-launches-40-MONTH-0-balance-transfer-credit-card.html}{https://www.thisismoney.co.uk}}
Key metrics such as income, market share, customer balances, and customer volumes from Virgin Money UK's annual reports are compared with a model run that seeks to emulate the launch of the credit card business and promotional card.
Although the model is calibrated on a time period later than the promotional card's launch, the calibrated parameters are expected to accurately reflect the conditions of the UK credit card market, given the relative stability of both periods.
This temporal discrepancy between calibration and validation is also observed in \citet{Carro_2022}, who conducted their time-wise validation on a period several years prior to their model's calibration date.

The time-wise validation model simulation comprises 30,000 customer agents over a period of $T$=100, with two lenders without promotions, and a third lender mimicking the behaviour of Virgin Money UK.
This lender initiates a credit card two years ($t$=24) into the simulation, and subsequently introduces a promotion on it three years later ($t$=60), mirroring the actions of Virgin Money UK.
This simulation was executed across five different seeds.
The characteristics of this lender in the model are plotted alongside those of Virgin Money UK, with both time series indexed against the value four years post the launch of the credit card business (the date when all properties of interest are available in Virgin Money UK reports).

\begin{figure}
     \centering
	\begin{subfigure}[t]{0.45\textwidth}
		\centering
		\includegraphics[width=1.0\textwidth]{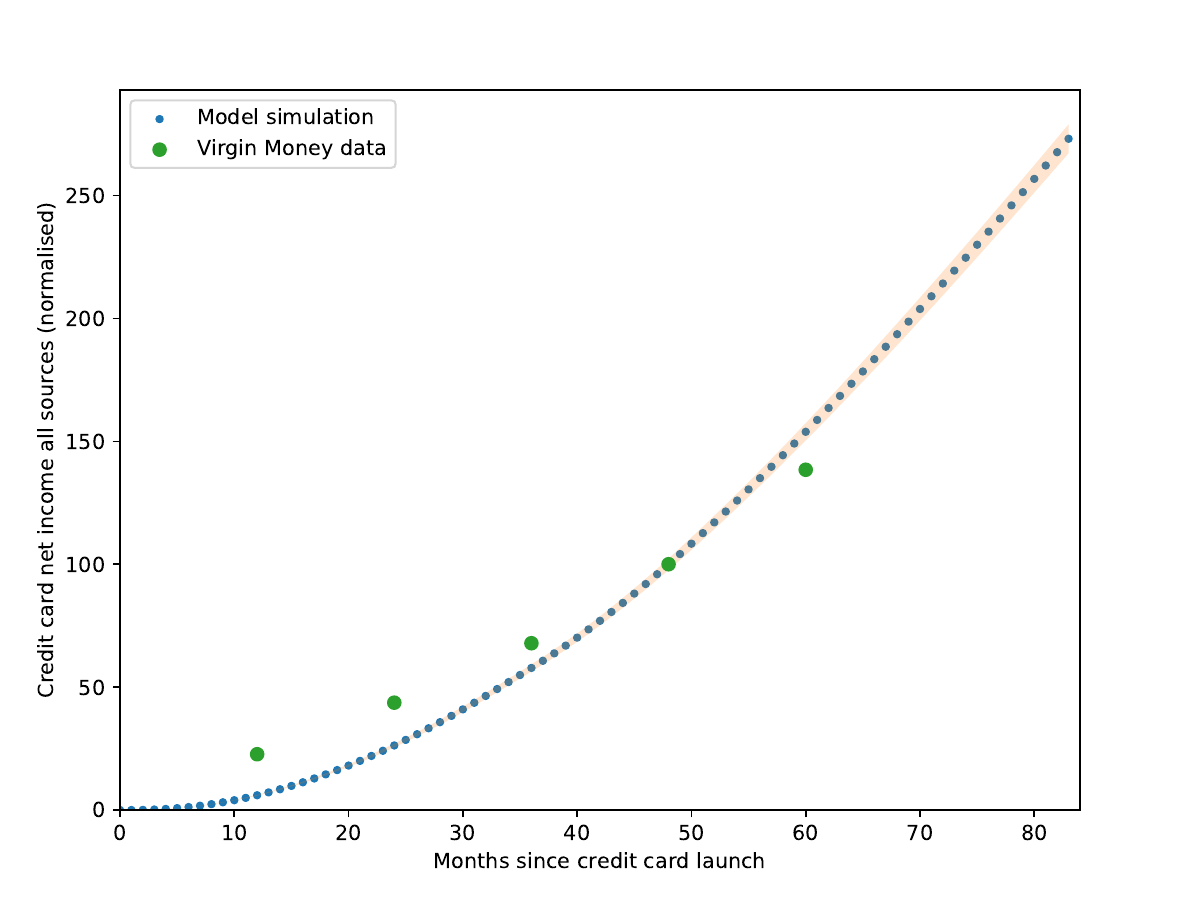}
		\caption{Historical credit card net income from interest for Virgin Money UK (green) and a simulated lender (blue) across time.}
		\label{fig:tw_val_Credit_card_net_income_from_interest_[£m]}
	\end{subfigure}
	\hfill
    \begin{subfigure}[t]{0.45\textwidth}
		\centering
		\includegraphics[width=1.0\textwidth]{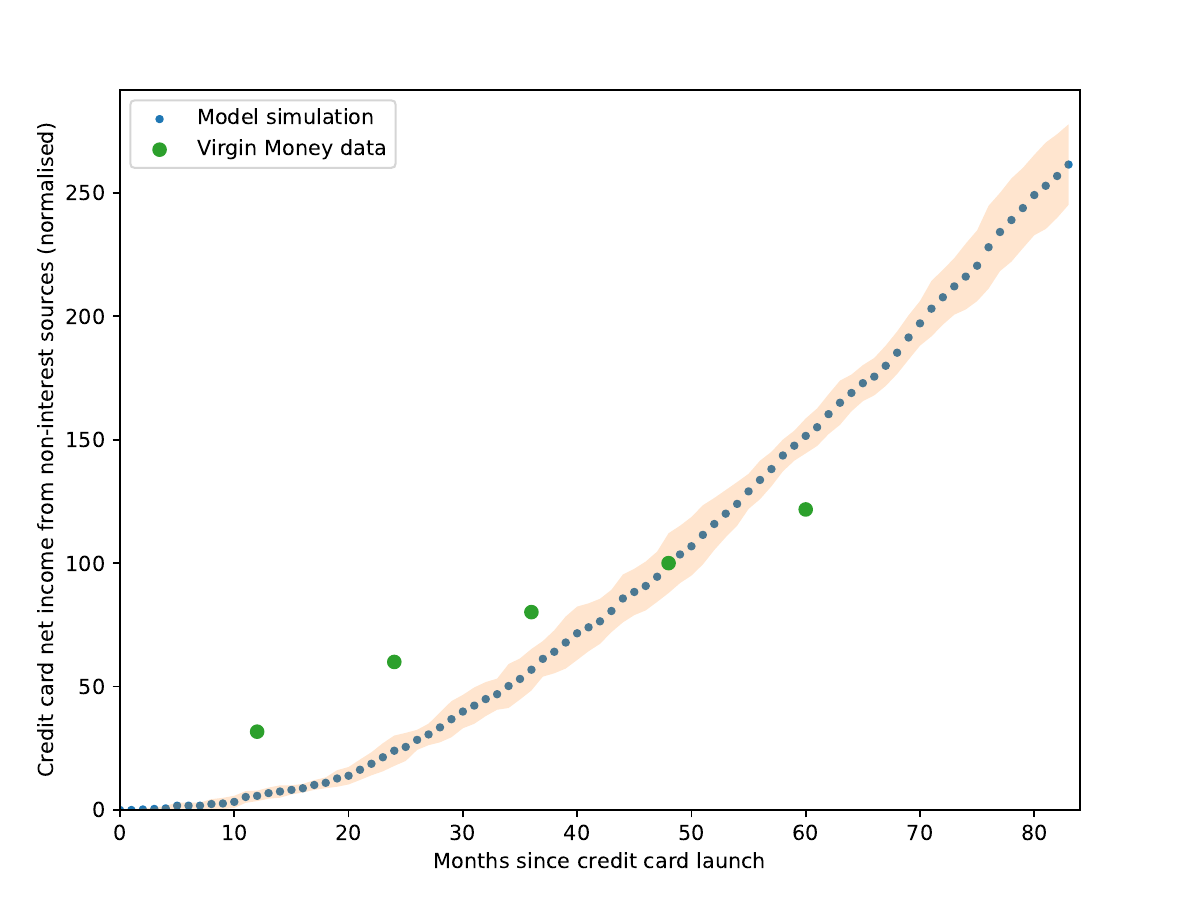}
		\caption{Historical credit card net income from non-interest sources for Virgin Money UK (green) and a simulated lender (blue) across time.}
		\label{fig:tw_val_Credit_card_net_income_from_non-interest_sources_[£m]}
	\end{subfigure}
	\par\medskip
	\begin{subfigure}[t]{0.45\textwidth}
		\centering
		\includegraphics[width=1.0\textwidth]{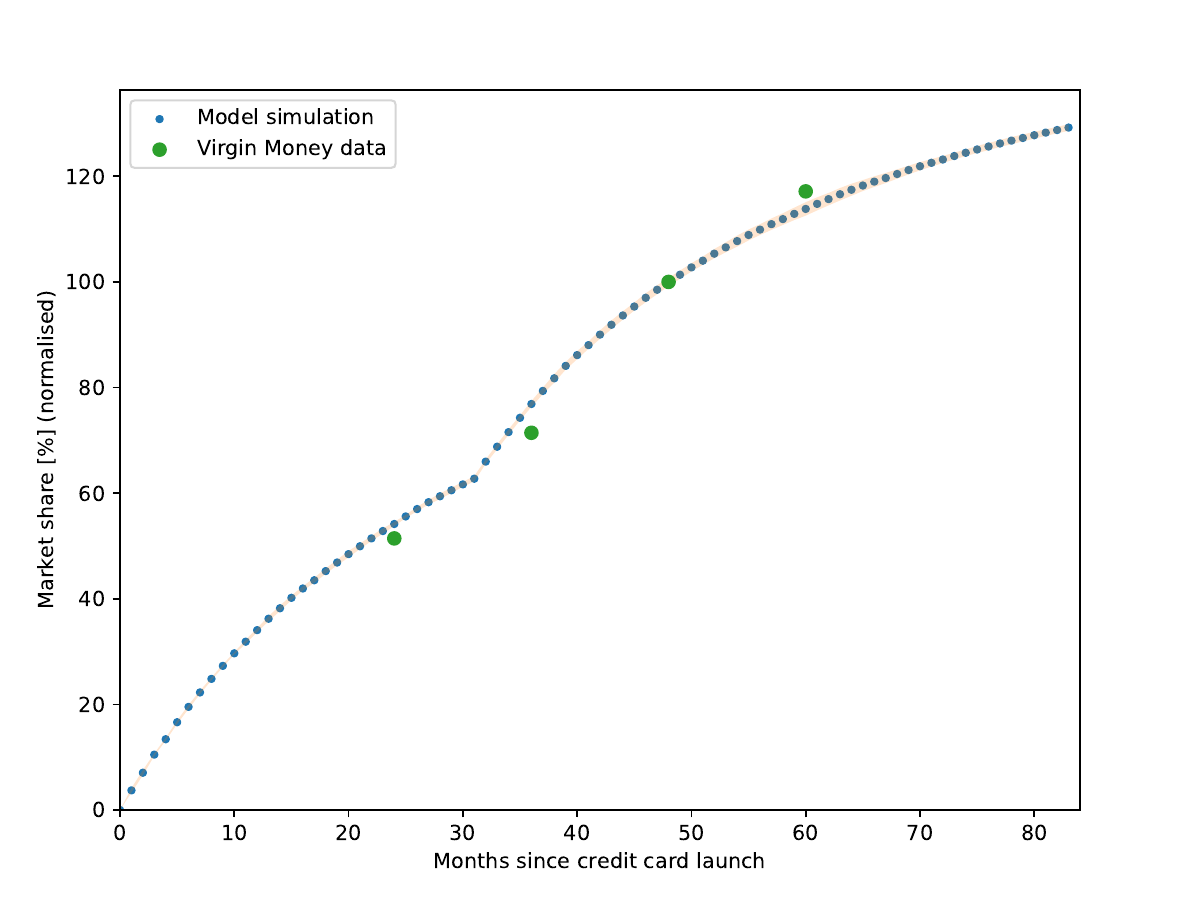}
		\caption{
		Historical market share for Virgin Money UK (green) and a simulated lender (blue) across time.}
		\label{fig:tw_val_Market_share}
	\end{subfigure}
	\hfill
	\begin{subfigure}[t]{0.45\textwidth}
		\centering
		\includegraphics[width=1.0\textwidth]{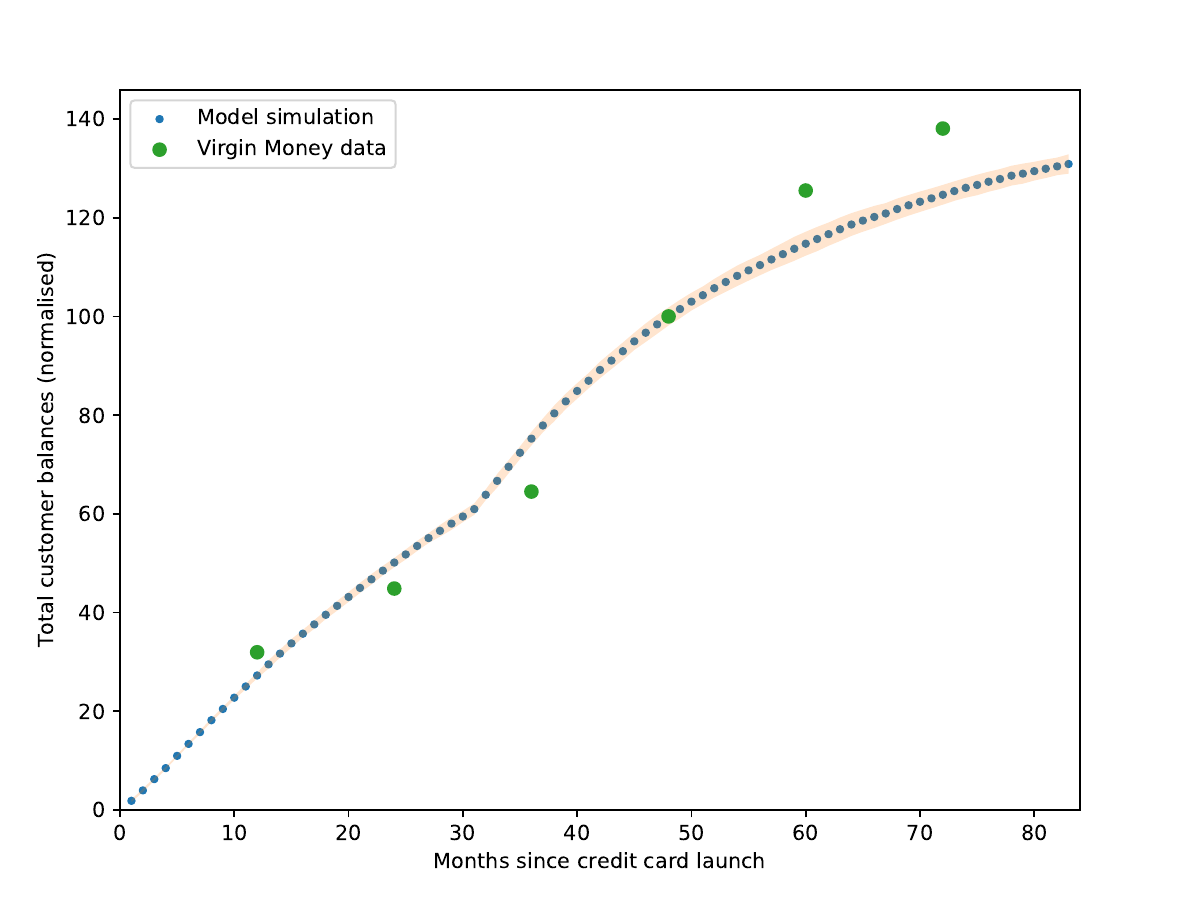}
		\caption{
		Historical total customer balances for Virgin Money UK (green) and a simulated lender (blue) across time.}
		\label{fig:tw_val_Total_customer_balances_[£b]}
	\end{subfigure}
     \caption{
     Comparison between historical values from Virgin Money UK (as reported in their annual reports) and the attributes of a simulated lender agent, designed to represent Virgin Money UK.
     The lender agent, following the actions of Virgin Money UK, introduces a credit card and launches a promotion on that card 30 months later.
     In all subplots, both time series are normalised to their value at 48 months (time steps) since the launch of the card.
     Uncertainty bands representing the standard deviation across 5 initial seeds are shown in orange.}
     \label{fig:qualitative_validate} 
\end{figure}

Figure \ref{fig:tw_val_Credit_card_net_income_from_interest_[£m]} illustrates the income derived from interest charges on accounts, as predicted by our model and as reported by Virgin Money UK.
The time-wise trends of both series exhibit a similar pattern of increasing profits year after year.
In Figure \ref{fig:tw_val_Credit_card_net_income_from_non-interest_sources_[£m]}, we compare the non-interest source profits predicted by our model with those reported by Virgin Money UK.
Despite some discrepancies, there is a rough agreement between the historical and simulated income from non-interest sources.
These discrepancies could be attributed to variations in the fees charged to customers in the model compared to real life, or to other income sources (e.g., annual card fees, income from cash back deals with partner companies) not being represented in the model.

In Figure \ref{fig:tw_val_Market_share}, we compare the market share reported by Virgin Money UK with the market share produced by our model.
Similarly, Figure \ref{fig:tw_val_Total_customer_balances_[£b]} compares the change in overall customer balance over time as reported by Virgin Money UK and as predicted by our model.
As can be observed, both the market share and the total customer balances are reasonably well-matched by the model's output, including the trend shift around the time of the promotion campaign's introduction.

In conclusion, our model demonstrates a reasonable agreement with historical trends across multiple variables.
The strong correlation in time-wise trends of consistently increasing profits from interest suggests that new credit cards accumulate profits in a manner similar to real-world scenarios.
The similar trends in the growth of total card balances and market share suggest that the model's underlying mechanisms effectively replicate customer behaviour.
The model's ability to match the trend of historical profits of a credit card provider lends it credibility as a tool for estimating optimal promotion strategies.

\FloatBarrier
\section{Experiments}\label{sec:experiments_label}

Following the calibration and validation of the model, we employ it to conduct a series of experiments to investigate the impact of various facets of a credit card promotion.

In all experiments, simulations are executed for a duration of $T$=120 (equivalent to 10 years), involving 3 lenders, with the parameters calibrated as displayed in Table \ref{tab:calibrated_parameters_table}.
We select a simulation size of 30,000 customer agents, a choice justified by the model stability this number of agents ensures, as demonstrated in Appendix \ref{subsec:experiment_results}, where 30,000 agents yield the anticipated market share for three identical lender agents.
Each experiment is replicated across five different initial seeds, also detailed in Appendix \ref{subsec:experiment_results}, to ensure model stability.
Unless specified otherwise, only the primary lender initiates promotions in the experiments discussed.
Promotions are not implemented during the initial two years (24 steps) of the simulations to allow for model stabilisation. 
Promotional campaigns are executed between steps 24 and 48 (corresponding to a promotion-availability window of 2 years), with the exception of the second experiment that examines the impact of varying promotion-availability windows and the fourth experiment that investigates the effects of promotions commencing at different time-wise points.
Consequently, if a lender offers a credit card with an 18-month promotion, the final step at which a customer could obtain a 0\% interest rate would be step 66.
Running the simulations for over 4 years post the closure of promotion-availability windows enables the observation of long-term profits derived from promotional offers.
The results of the experiments for each simulation configuration are reported as the mean and sample standard deviation across five initial different seeds.

\subsection{Impact of credit card promotion interest-free duration}\label{sec:rq1_sec}

To investigate the influence of varying interest-free durations, we conducted six simulations where two providers offered no promotions, while the primary provider offered an interest-free period of 0 (no promotion), 6, 12, 24, 36, and 48 months between steps $t$=24 and $t$=48.
In all simulations, both the primary and competing lenders applied a 20\% retail APR after the introductory promotions on cards concluded. 
The complete results of this experiment are presented in Table \ref{tab:RQ1-table} in Appendix \ref{subsec:experiment_results}.

Figure \ref{fig:Research_question_1_lender_number_applications} illustrates the impact of introducing credit card promotions on the number of approved credit card applications.
Lenders offering promotions witness a surge in accepted applications between steps 24 and 48, during the promotional campaign, suggesting that promotional cards are considerably more attractive to customers than non-promotional cards. 
The two longest promotions initially receive a higher number of accepted applications due to the increased appeal of the extended interest-free period. 
However, the number of applications eventually declines to similar levels as the other interest-free durations, implying that promotions must be sustained over extended periods for market share to continue growing or remain stable.  

Longer promotions also enable lenders to secure a higher market share by the end of the promotion-availability window, as depicted in Figure \ref{fig:Research_question_1_lender_market_share}. 
As anticipated, the primary lender offering no promotion maintains a market share of approximately one-third throughout the simulation.
Conversely, lenders offering promotions experience a substantial increase in their market share during their promotional campaigns, followed by a gradual decrease.
We observe that longer interest-free durations result in larger market shares, even several years post-campaign, suggesting that promotions enhance market share over an extended duration (in the absence of competing offers), and that the two longest promotions achieve very similar market shares.

In terms of customer behaviour, Figure \ref{fig:Research_question_1_missed_payments} reveals that the proportion of missed payments for a lender decreases when they offer a more generous promotion.
This reduction in risk when lenders offer better promotions aligns with the adverse selection phenomenon observed in the credit card market by \cite{Agarwal_2010}, who discovered that individuals responding to less favourable promotion offers had poorer credit qualities than those responding to more generous credit offers.

As demonstrated in Figure \ref{fig:Research_question_1_lender_profit}, the length of the interest-free duration significantly impacts the discounted profit for lenders at the simulation's end. 
The primary lender offering the 48-month promotion records the lowest profit, followed by the lender with no promotion, while the lender with the highest profit at the final step 120 is the one that offered a 12-month promotion, closely followed by the lender with a 6-month promotion. 

Despite the primary lender offering the longest interest-free duration of 48 months amassing the largest market share, this does not correspond to the highest profits in the simulations. 
The profits for this lender do not initially grow as rapidly as those of other lenders offering promotions, even earning less than the zero-promotion lender during the simulation. 
However, by the final step of the simulation, the profit of this lender is increasing at a rapid pace.

Our model suggests that launching a promotional credit card is generally beneficial for lenders.
Our experiment indicates that the most effective promotion, in terms of profit from interest, over the simulation's time frame (8 years post-promotion launch) is achieved with an interest-free duration of 12 months, with the largest market share secured by the lender offering a 48-month promotion.
However, the maximum profit around step 50 in the model is actually achieved by the lender offering a short promotion of 6 months, suggesting that lenders should select their interest-free duration based on their specific objectives.

\begin{figure}
     \centering
     \begin{subfigure}[t]{0.45\textwidth}
	 	\centering
		\includegraphics[width=1.0\textwidth]{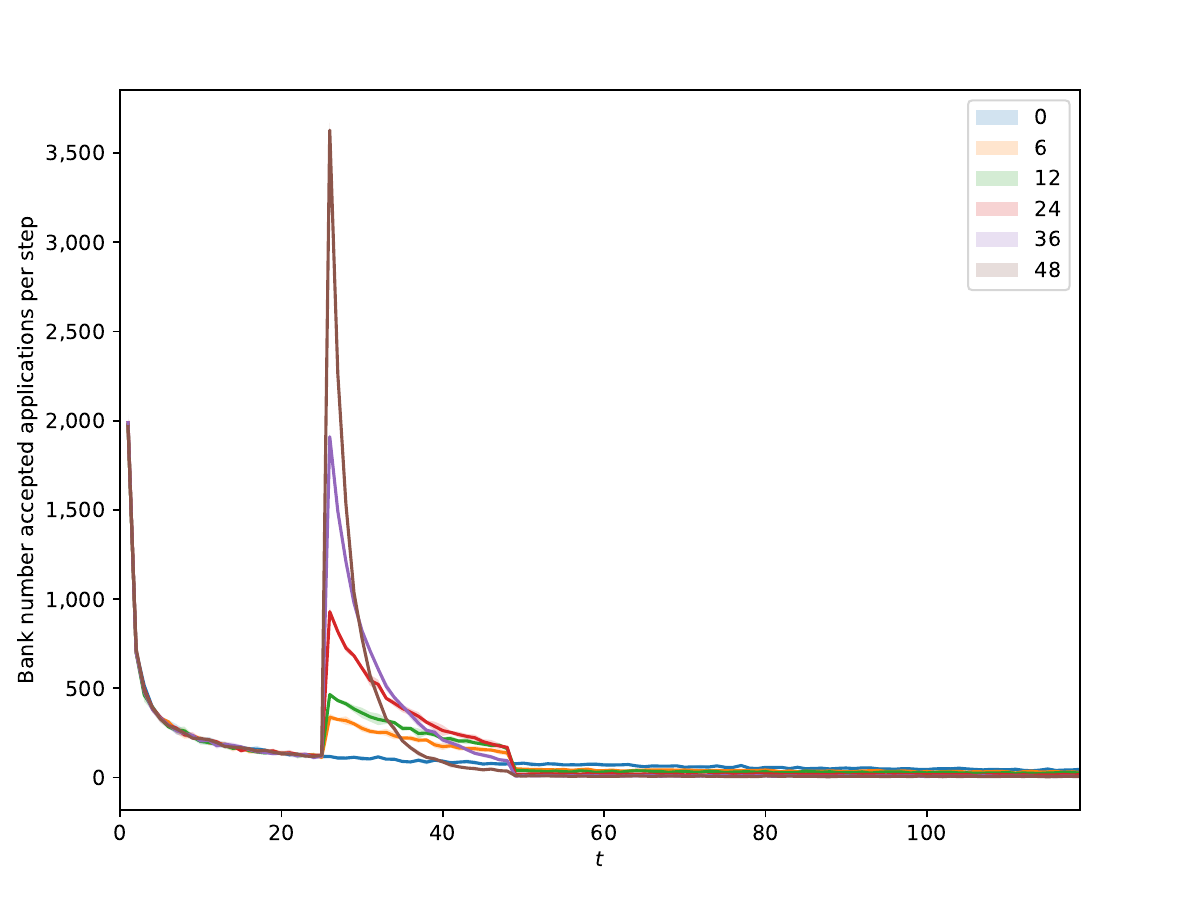} 
		\caption{Number of accepted credit card applications for the primary lender with different interest-free durations.}
		\label{fig:Research_question_1_lender_number_applications}
     \end{subfigure}
     \hfill
     \begin{subfigure}[t]{0.45\textwidth}
     	\centering
		\includegraphics[width=1.0\textwidth]{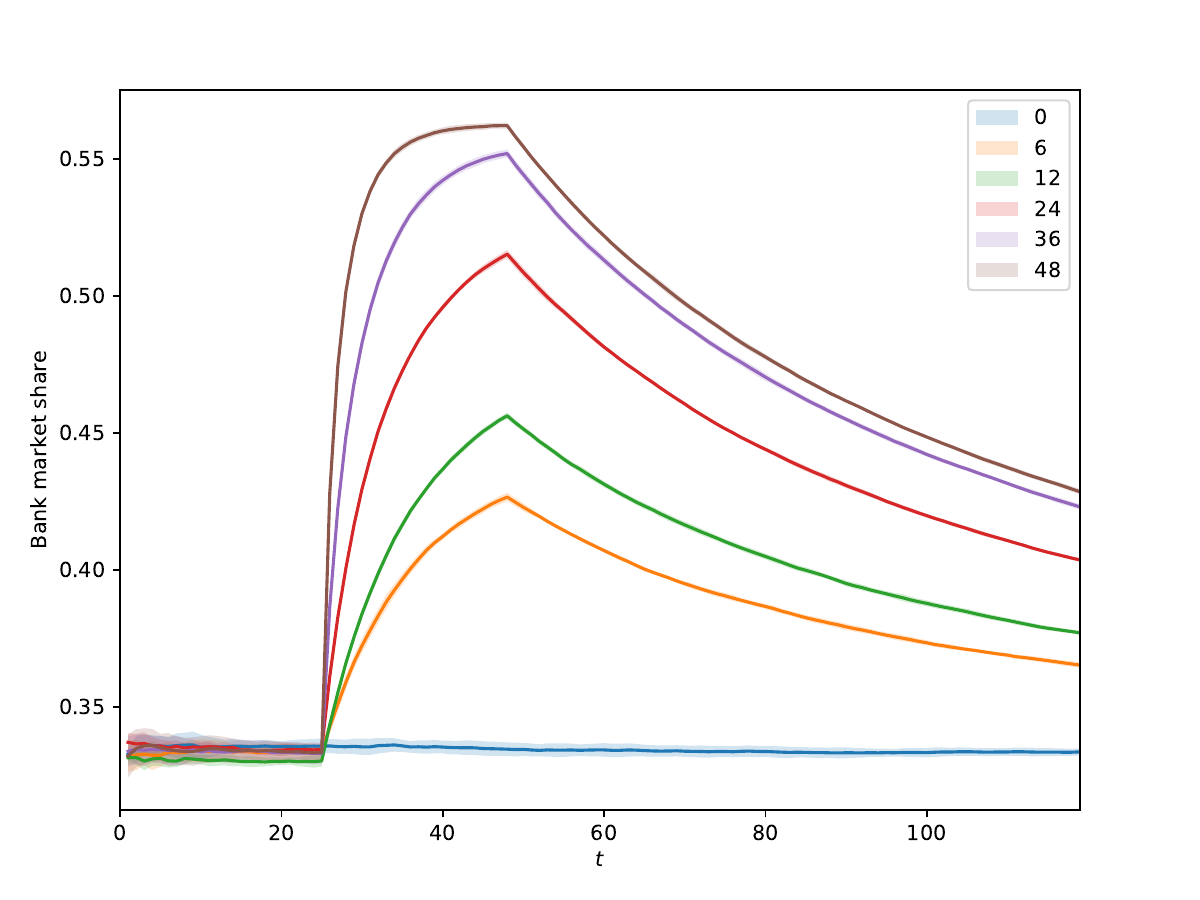}
		\caption{Proportion market share for the primary lender with different interest-free durations.}
		\label{fig:Research_question_1_lender_market_share}
     \end{subfigure}
     \par\medskip
     \begin{subfigure}[t]{0.45\textwidth}
     	\centering
		\includegraphics[width=1.0\textwidth]{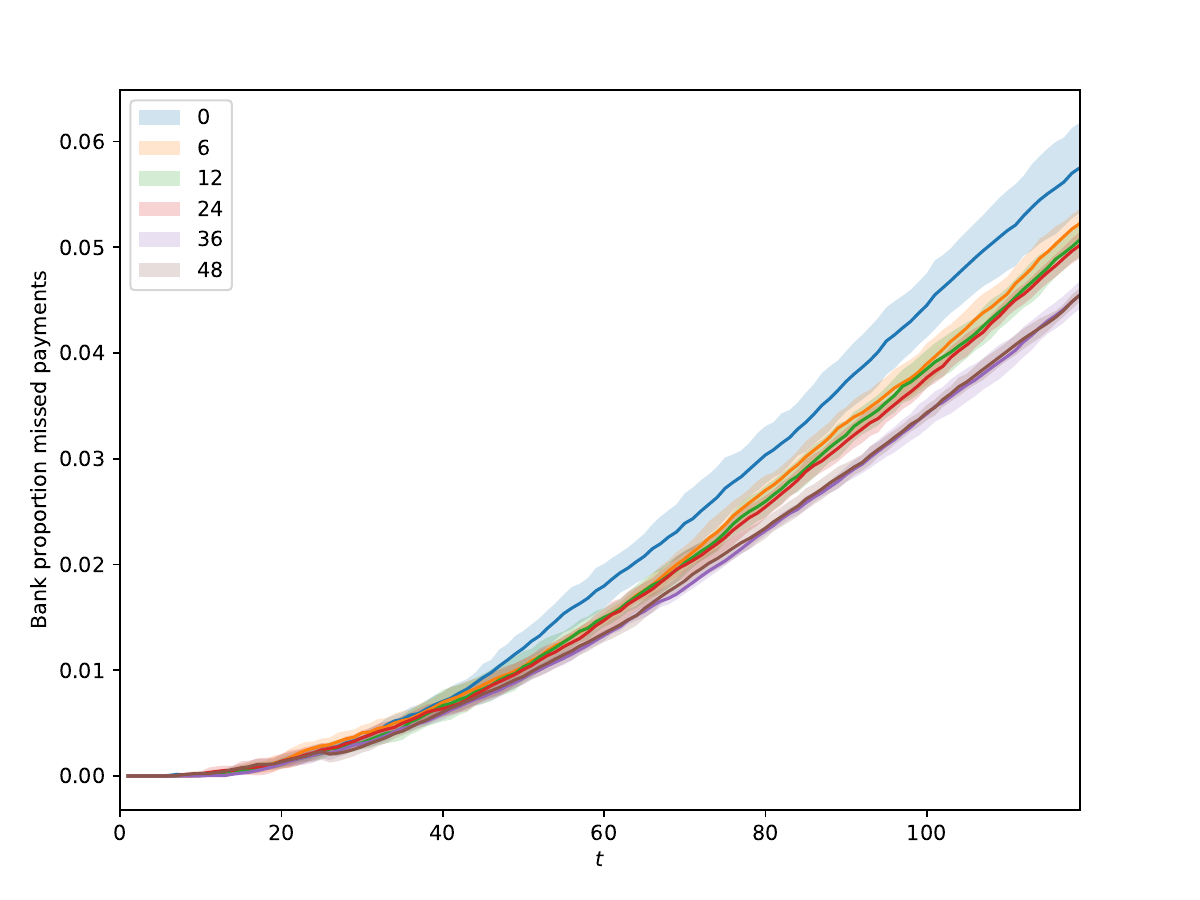}
		\caption{Proportion of cumulative missed payments for the primary lender with different interest-free durations.}
		\label{fig:Research_question_1_missed_payments}
     \end{subfigure}
	 \hfill     
     \begin{subfigure}[t]{0.45\textwidth}
     	\centering
		\includegraphics[width=1.0\textwidth]{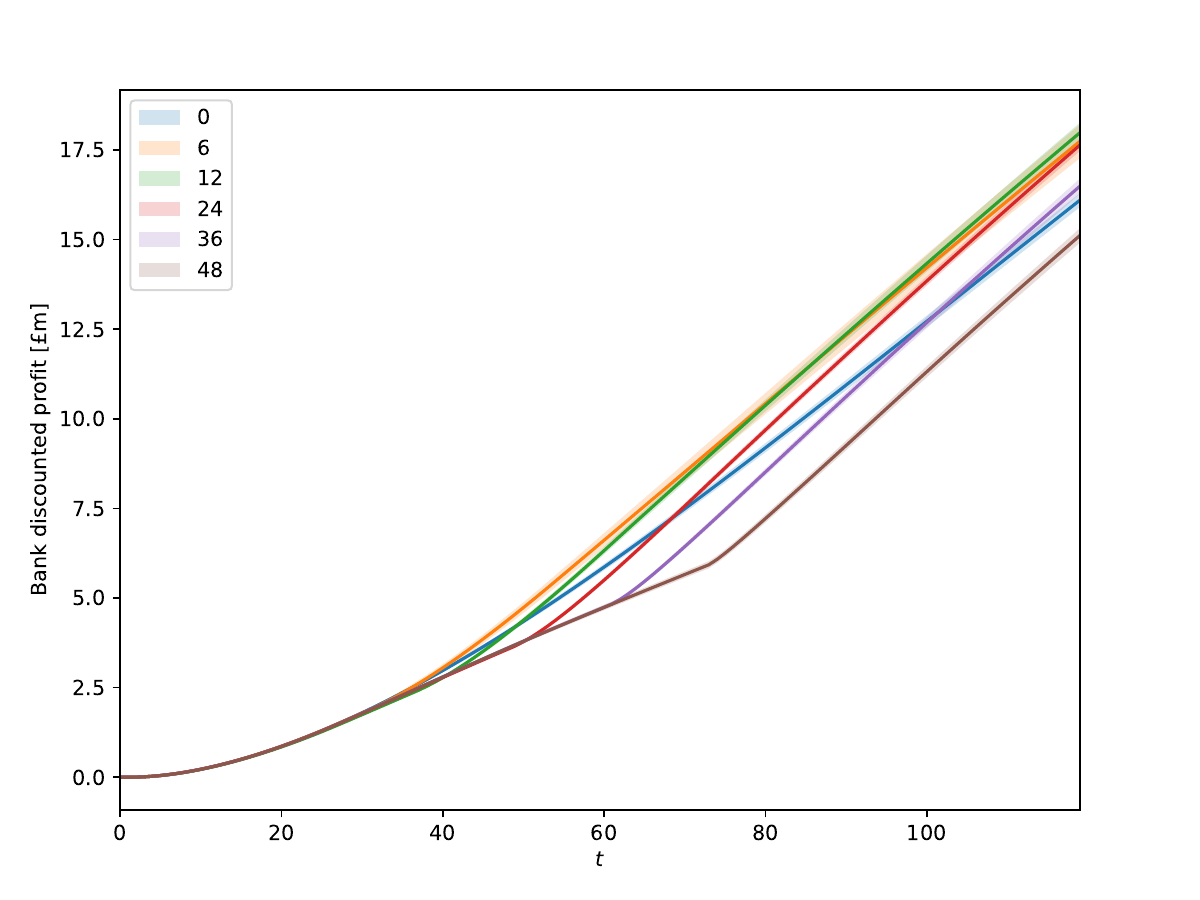}
		\caption{Cumulative discounted profit for the primary lender with different interest-free durations.}
		\label{fig:Research_question_1_lender_profit}
     \end{subfigure}
     \caption{Experimental results for the primary lender offering 0- (no promotion), 6-, 12-, 24-, 36- and 48-month interest-free promotions.
     Solid lines represent the mean and shaded areas correspond to one standard deviation.}
     \label{fig:Research_question_1}
\end{figure}

\FloatBarrier

\begin{table}[ht]
\centering
\begin{threeparttable}
\caption{Average attributes of customer agents for different interest-free durations run by the primary lender agent.
Values are across an experiment lasting 10 years and uncertainties presented are the sample standard deviations across five seeds.}
\label{tab:RQ1-customers}
\centering
\begin{tabular}{rrrrrrr}
\toprule
\begin{tabular}[l]{@{}l@{}}Interest-free\\ duration (months) \end{tabular} & \begin{tabular}[r]{@{}r@{}}Average total\\ balance (\pounds) \end{tabular} & \begin{tabular}[r]{@{}r@{}}Average total\\ interest paid (\pounds) \end{tabular} & \begin{tabular}[r]{@{}r@{}}Average total\\ usage amount (\pounds)\end{tabular} & \begin{tabular}[r]{@{}r@{}}Average total\\ credit limit (\pounds)\end{tabular} & \begin{tabular}[r]{@{}r@{}}Proportion accounts\\ with missed payment\end{tabular} \\
\midrule
                                0 &             3,846 ± 23 &                   2,980 ± 30 &               163,350 ± 180 &                  7,250 ± 30 &                                               0.0209 ± 0.0016 \\
                                6 &             3,840 ± 50 &                   2,960 ± 60 &               163,340 ± 260 &                  7,300 ± 40 &                                                 0.0203 ± 0.0005 \\
                               12 &             3,880 ± 40 &                   2,990 ± 50 &               163,700 ± 300 &                  7,420 ± 40 &                                                 0.0203 ± 0.0007 \\
                               24 &             3,908 ± 28 &                   2,930 ± 40 &               163,680 ± 270 &                  7,740 ± 40 &                                                 0.0214 ± 0.0007 \\
                               36 &             3,910 ± 40 &                   2,840 ± 50 &               163,600 ± 300 &                  7,950 ± 50 &                                               0.0203 ± 0.0006 \\
                               48 &             3,909 ± 23 &                   2,740 ± 40 &               163,620 ± 140 &                  7,989 ± 26 &                                             0.0205 ± 0.0027 \\
\bottomrule
\end{tabular}

\end{threeparttable}
\end{table}

We further analysed customer attributes across different interest-free durations, with the results presented in Table \ref{tab:RQ1-customers}.
Interestingly, while the average total balance and average total amount used on the credit cards appear to increase with the length of the interest-free duration offered by a lender, the average total interest paid by customers decreases.
This can be explained by customers taking advantage of the interest-free periods on the cards to increase their card balances and not pay interest for a period of time. 
The larger lines of credit available to customers as interest-free duration increases suggest that these promotions may enhance access to credit, without necessarily increasing the amount spent on interest on cards, but could potentially escalate customer indebtedness.

Despite the increase in total balances on customer credit cards with more generous promotions, the fractions of accounts with missed payments remain approximately constant, indicating no significant increase in the risk of default.
As previously discussed in the analysis of the impact of interest-free durations on the lender, the proportion of missed payments on the accounts they open decreases as the promotion period lengthens.
Given that the number of accounts with missed payments appears to remain constant regardless of the interest-free duration offered, this implies that in a market where a lender is offering a promotion, lenders who are not offering a promotion may be assuming riskier accounts.

\FloatBarrier

\subsection{Impact of credit card promotion-availability windows}

To examine the effect of varying promotion-availability windows (i.e., the period that a lender runs a promotion for), we conducted six simulations where the primary lender offered a 12-month interest-free promotion for a total period of 0 (no promotion), 6, 12, 24, 36, or 48 months, starting at $t$=24.
The complete results of these simulations are presented in Table \ref{tab:RQ2-table} in Appendix \ref{subsec:experiment_results}.

\begin{figure}
     \centering
     \begin{subfigure}[t]{0.45\textwidth}
		\centering
		\includegraphics[width=1.0\textwidth]{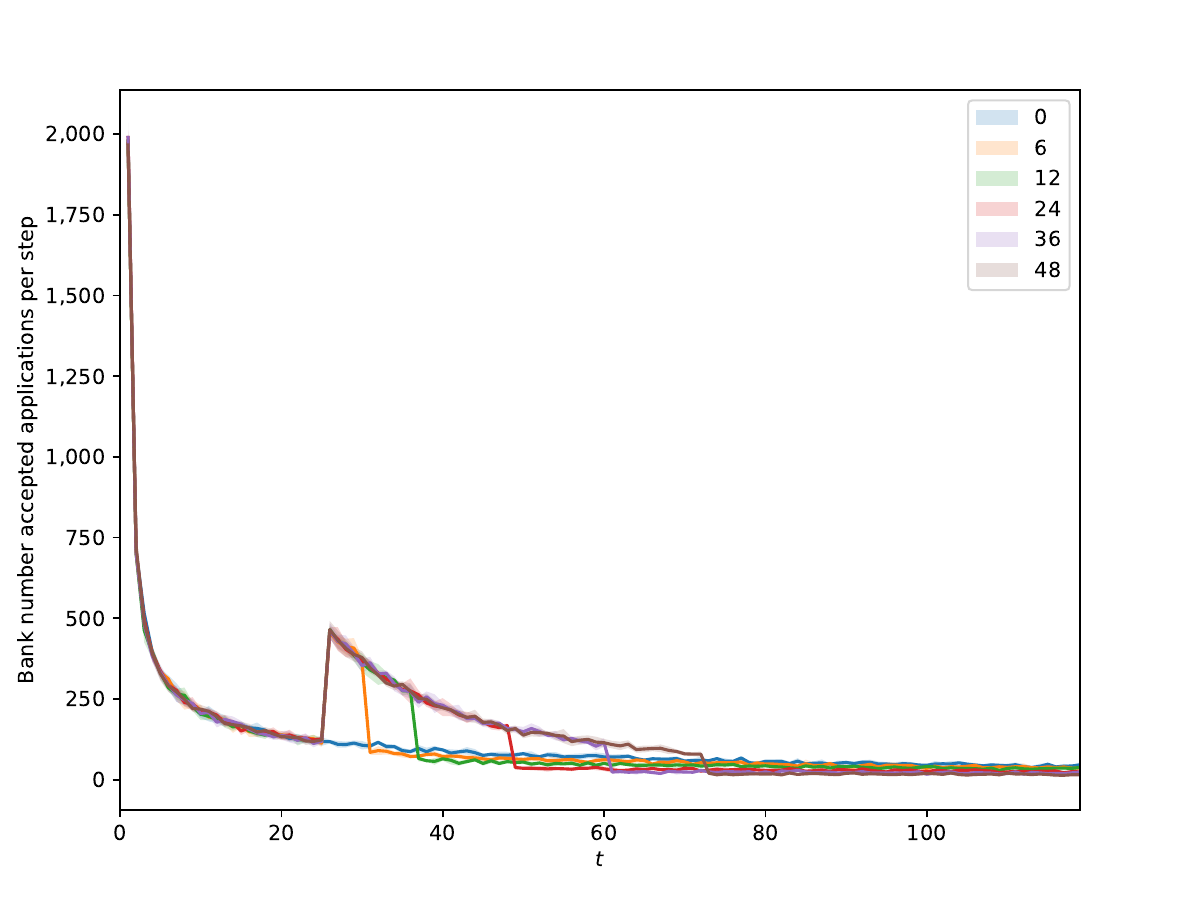}
		\caption{Number of accepted applications for the primary lender with varying promotion-availability windows.}		
		\label{fig:Research_question_2_lender_number_applications}
     \end{subfigure}
     \hfill
     \begin{subfigure}[t]{0.45\textwidth}
		\centering		
		\includegraphics[width=1.0\textwidth]{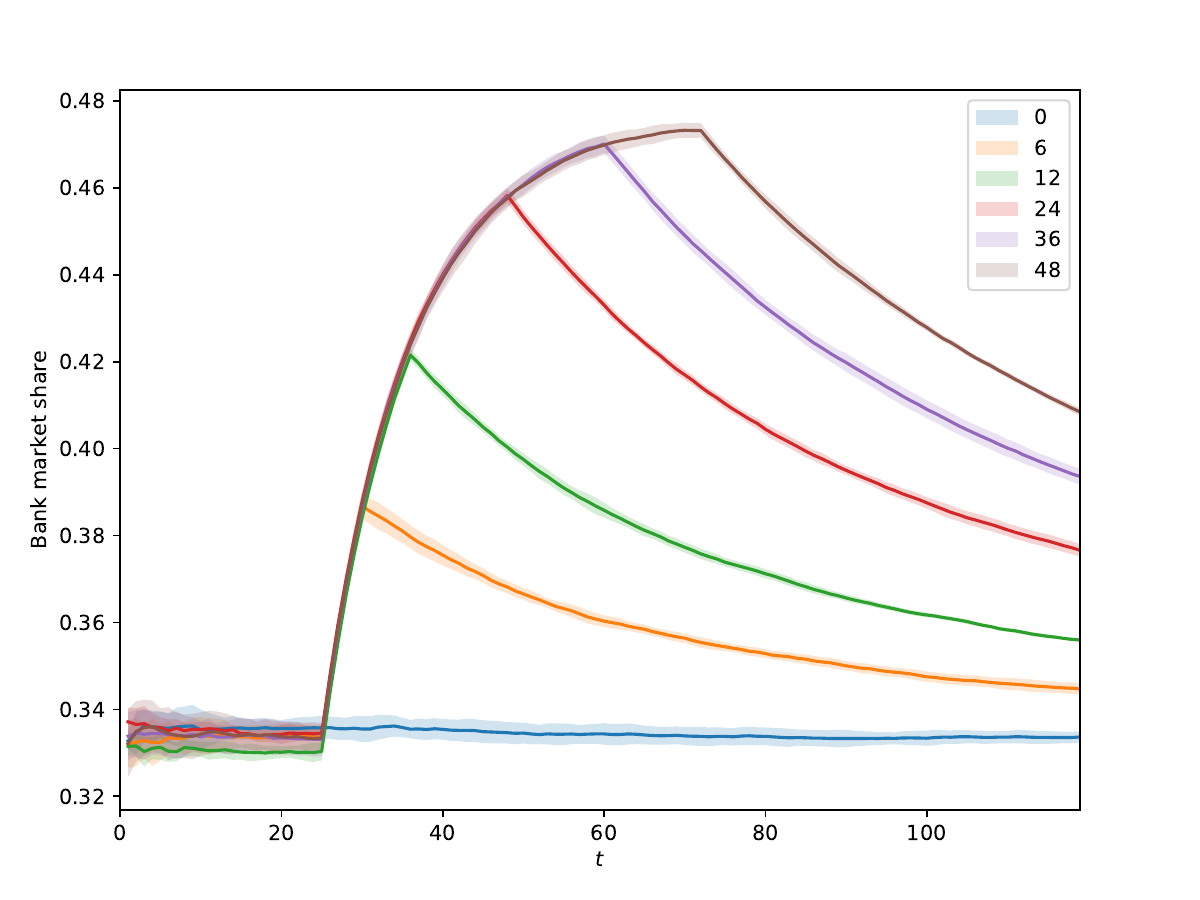}
		\caption{Proportion market share for the primary lender with varying promotion-availability windows}
		\label{fig:Research_question_2_lender_market_share}
     \end{subfigure}
     \par\medskip
     \begin{subfigure}[t]{0.45\textwidth}
		\centering		
		\includegraphics[width=1.0\textwidth]{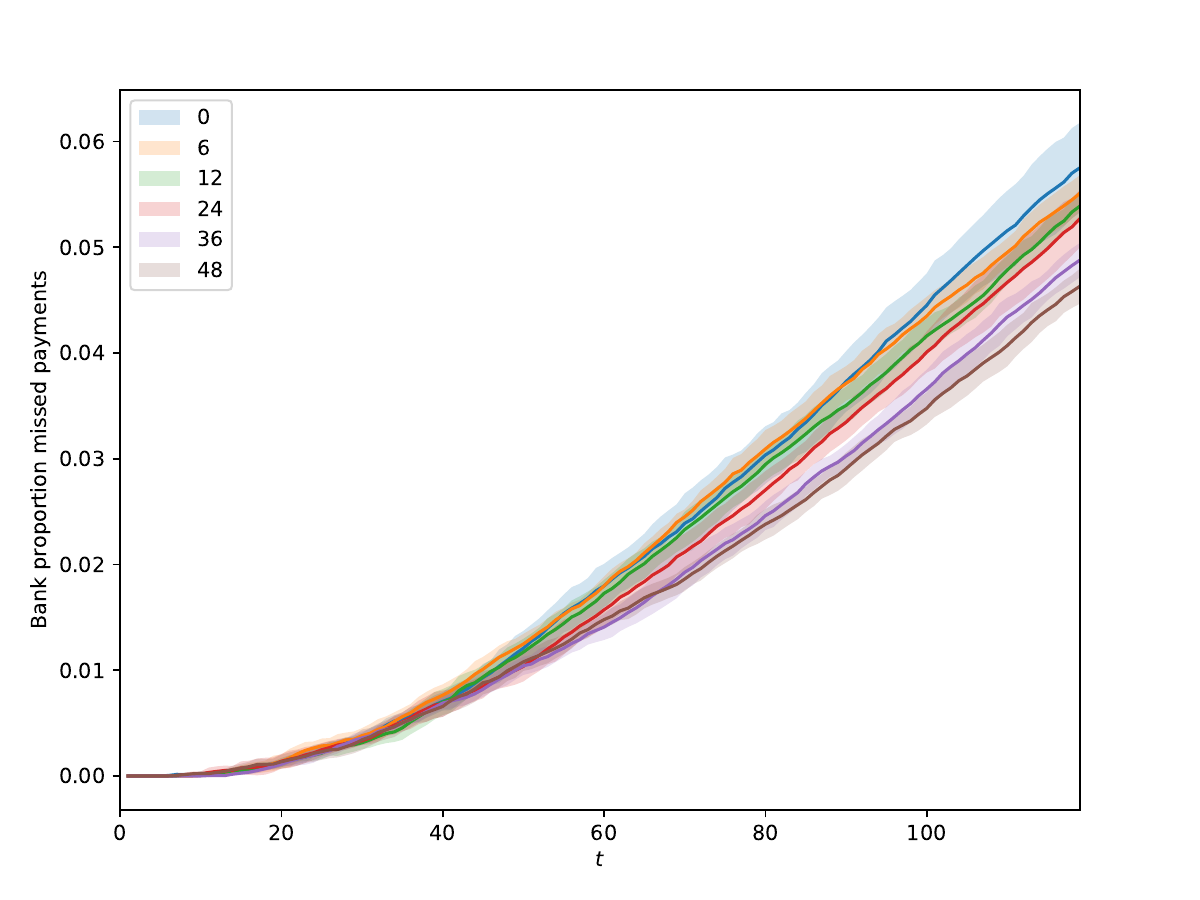}
		\caption{Proportion of cumulative missed payments for the primary lender with varying promotion-availability windows.}
		\label{fig:Research_question_2_missed_payments}
     \end{subfigure}
	 \hfill
     \begin{subfigure}[t]{0.45\textwidth}
		\centering		
		\includegraphics[width=1.0\textwidth]{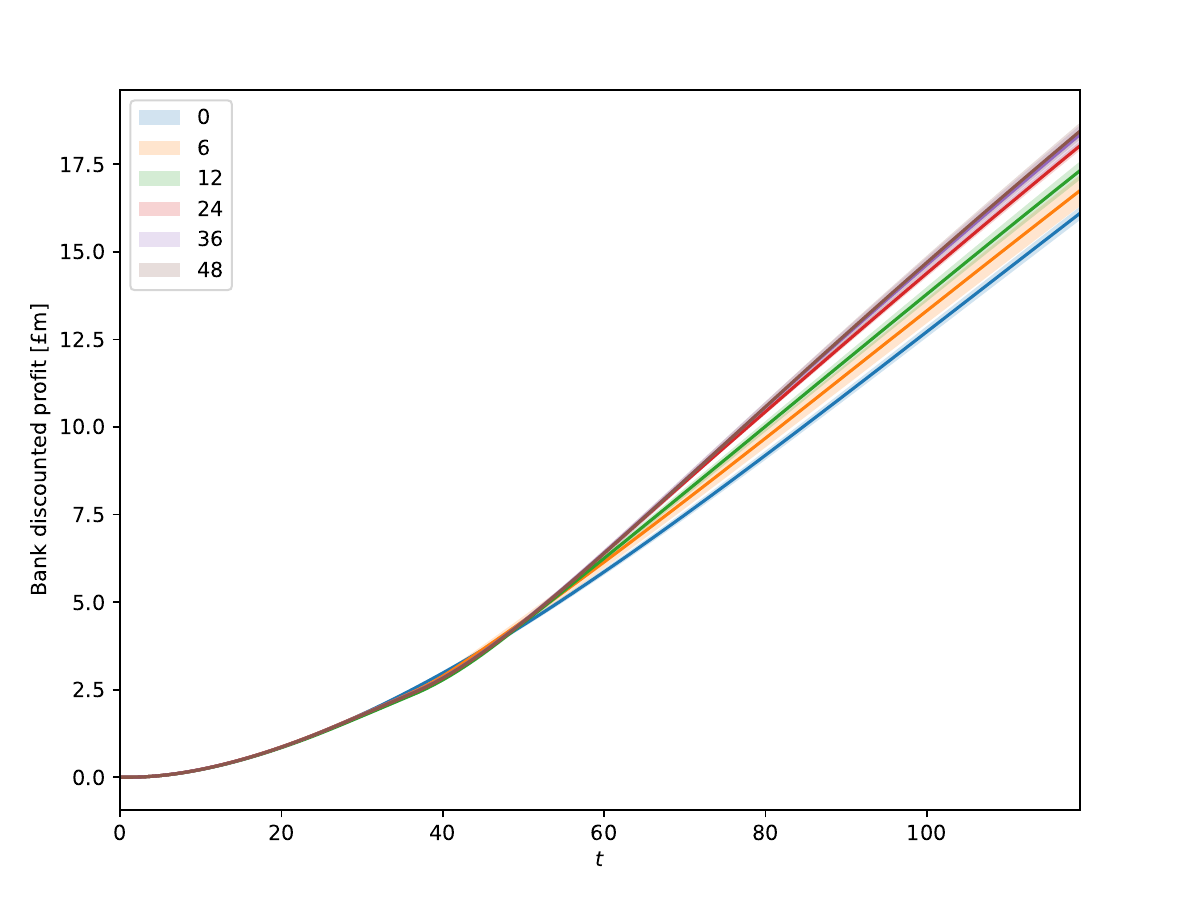}
		\caption{Cumulative discounted profit for the primary lender with varying promotion-availability windows.}
		\label{fig:Research_question_2_lender_profit}
     \end{subfigure}
     \caption{Experimental results for the primary lender offering 6-, 12-, 18-, 24-, 36- and 48-month promotion-availability windows.
     Solid lines represent the mean and shaded areas correspond to one standard deviation.}
     \label{fig:Research_question_2}
\end{figure}

As depicted in Figure \ref{fig:Research_question_2_lender_number_applications}, the number of approved credit card applications increases for the duration of the promotional campaign. 
Moreover, the simulation results indicate that the market share of the lender offering the promotion expands during the campaign, with a longer promotion-availability window resulting in a higher market share by the end of the simulation, as shown in Figure \ref{fig:Research_question_2_lender_market_share}.
Similar to the previous experiment on interest-free duration, the market share begins to decline once the campaign concludes, yet the lender retains a larger market share than its competitors for several time steps post-campaign.

In terms of profitability, Figure \ref{fig:Research_question_2_lender_profit} demonstrates that a longer promotion-availability window is generally beneficial. 
However, the profit difference between a 3-year and a 4-year promotion-availability window is marginal. 
If the profits projected over a 10-year horizon for both a 3-year and a 4-year promotion-availability window are similar, a lender might prefer the 4-year window due to the larger market share it garners.
Figure \ref{fig:Research_question_2_missed_payments} also suggests that longer promotion-availability windows correlate with a slight decrease in the proportion of missed payments on accounts, implying that extending the promotion-availability window does not increase the lender's risk.
The observation that longer promotion-availability windows lead to increased profit without a corresponding increase in risk aligns with the common practice among many UK lenders of running credit card promotional campaigns for extended periods.

\FloatBarrier

\subsection{Impact of competing credit card promotions}

To assess the optimal strategy in a competitive environment, we conducted additional simulations where one lender offered no promotion, a competing lender offered a promotion of 6, 12, 18, or 24 months, and the primary lender offered a promotion of 6, 12, 18, or 24 months. 
This resulted in 16 distinct simulation combinations. 
For both the primary and competitor lender agents, these promotions were implemented from $t$=24 to $t$=48. 
The complete results from this experiment are presented in Table \ref{tab:RQ3-table} in Appendix \ref{subsec:experiment_results}.

Figure \ref{fig:Research_question_3_competitor_6_months_profit} displays the results of the simulations where the competing lender offers a 6-month promotion. 
It is evident that when both the competitor and the primary lender offer the same interest-free duration of 6 months, the profit is comparable. 
To enhance profits, the primary lender needs to offer a longer interest-free duration than the competitor. 
In the short term, a 12-month promotion yields the most significant difference in profit. 
However, by the end of the simulation, longer promotions have further amplified this difference in profit between the lenders. 
This suggests that lenders should consider their short and long-term objectives when deciding how to respond to a competing promotion.

This trend of needing to provide a longer promotion than a competitor to maximise profits is consistent across the other simulations, where a competitor offered longer promotions of 12, 18, and 24 months, as depicted in Figure \ref{fig:Research_question_3_competitor_12_months_profit},  \ref{fig:Research_question_3_competitor_18_months_profit},  \ref{fig:Research_question_3_competitor_24_months_profit}. 
In all simulations, if the primary lender matched the competitor promotion, there is a minimal difference in profit. 
Generally, a promotion that is slightly better than a competitor's will yield a greater profit in the short term, however, significantly better promotions may increase long-term profits.
Whenever a competitor offers a better promotion than the primary agent, the greater the difference in promotions, the larger the difference in profits, as can be clearly observed in Figures \ref{fig:Research_question_3_competitor_18_months_profit} and \ref{fig:Research_question_3_competitor_24_months_profit}.
Table \ref{tab:RQ3_profit_percentage_difference_table} summarises the percentage profit difference between the two lenders.

This experiment suggests that, generally, the lender offering a longer interest-free promotion will have higher profits.
Moreover, for each of the promotions offered by the competitor, the primary lender gained the largest profit at the end of the simulation by offering the longest possible interest-free duration (although the uncertainties of these values overlap with the profits from other interest-free durations), as shown in Table \ref{tab:RQ3_profit_percentage_difference_table}.
These results therefore suggest that the optimal strategy to maximise profit in the context of a competing promotion is to offer the most attractive promotion possible, outperforming the competitor, while offering a promotion that is inferior to a competitor's will lead to a significant loss of profit. 
However, similar to the trend discussed in Section \ref{sec:rq1_sec}, short-term profit is sometimes best maximised by using shorter interest-free durations (as long as they are superior to the competition), once again highlighting that the optimal action for a lender depends on their goals and their competitors.
In contrast to the optimal interest-free duration of 12 months found in Section \ref{sec:rq1_sec}, this experiment underscores that the optimal interest-free duration of a promotion varies dramatically with the competitor's promotion, emphasising the importance for lenders to consider the competition when pricing a promotion.

\begin{figure}
	 \centering
     \begin{subfigure}[t]{0.45\textwidth}
		\centering
		\includegraphics[width=1.0\textwidth]{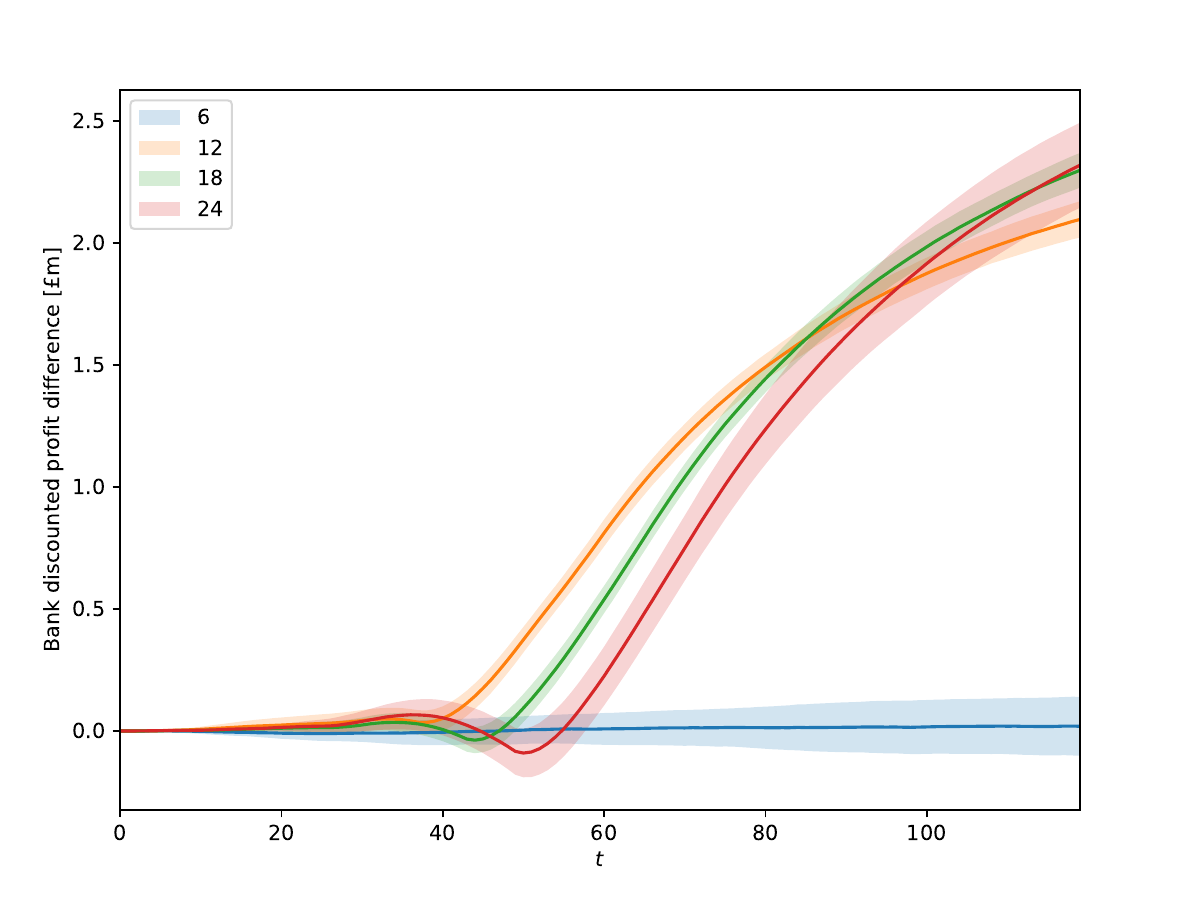}
		\caption{Difference in discounted profit of primary lender compared to a competitor offering a 6-month promotion.}
		\label{fig:Research_question_3_competitor_6_months_profit}
     \end{subfigure}
       \hfill
     \begin{subfigure}[t]{0.45\textwidth}
		\centering
		\includegraphics[width=1.0\textwidth]{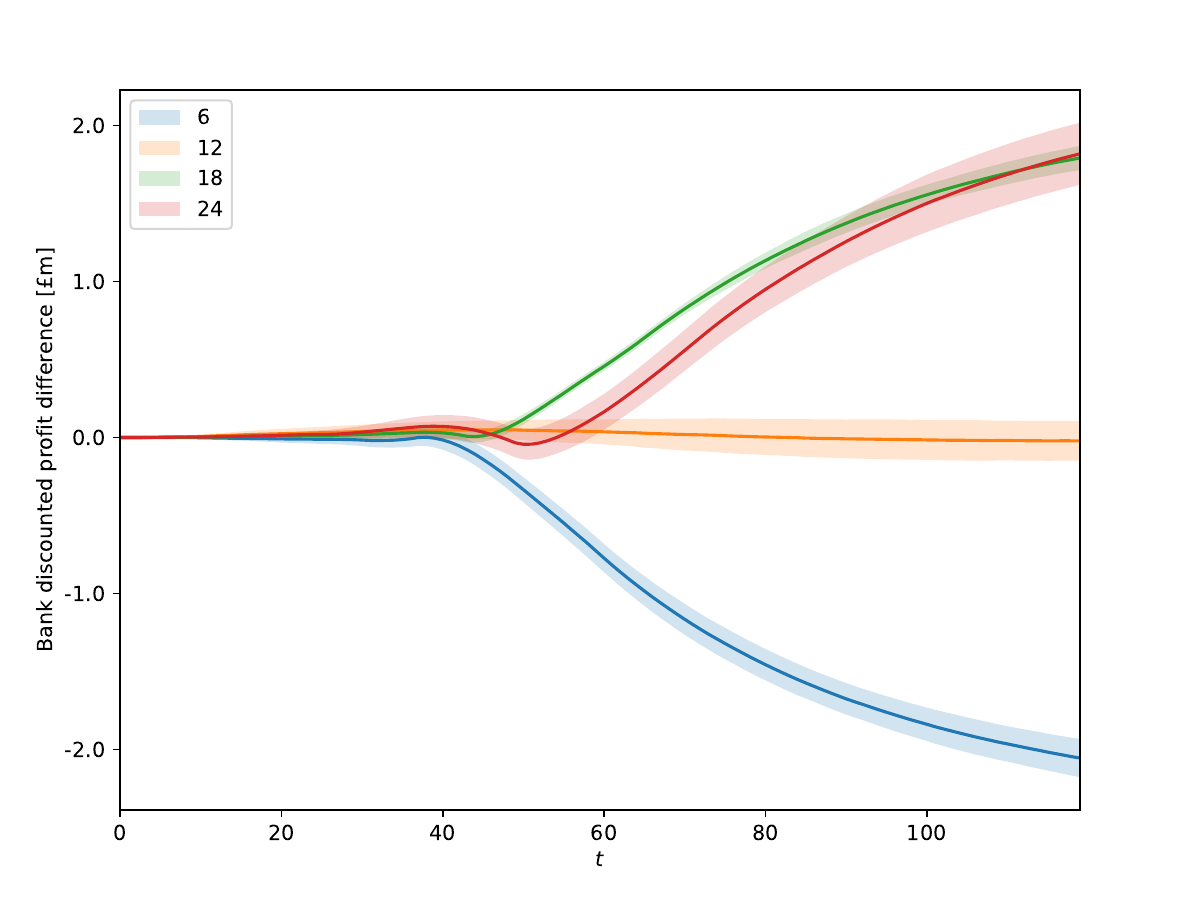}
		\caption{Difference in discounted profit of primary lender compared to a competitor offering a 12-month promotion.}
		\label{fig:Research_question_3_competitor_12_months_profit}
     \end{subfigure}
     \par\medskip
     \begin{subfigure}[t]{0.45\textwidth}
		\centering
		\includegraphics[width=1.0\textwidth]{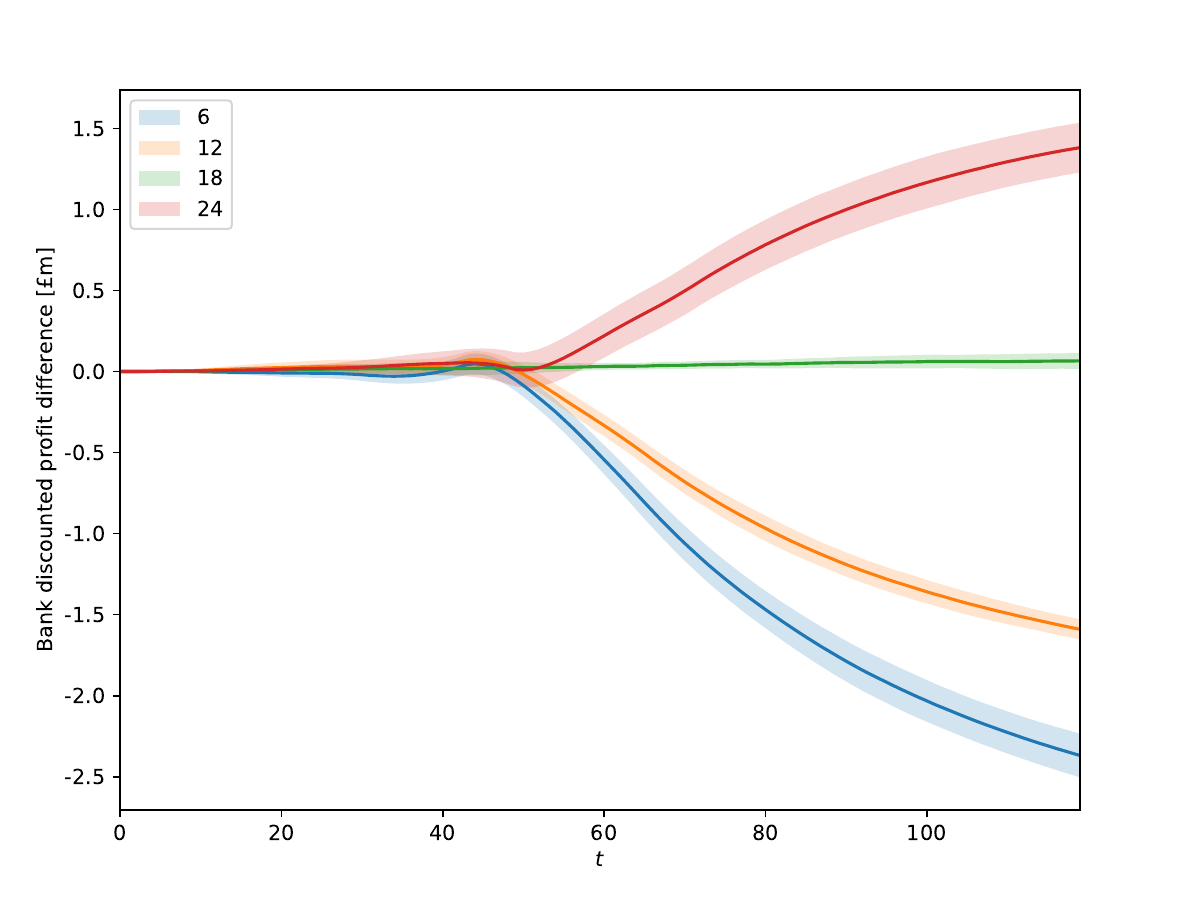}
		\caption{Difference in discounted profit of primary lender compared to a competitor offering a 18-month promotion.}
		\label{fig:Research_question_3_competitor_18_months_profit}
     \end{subfigure}
     \hfill
     \begin{subfigure}[t]{0.45\textwidth}
		\centering
		\includegraphics[width=1.0\textwidth]{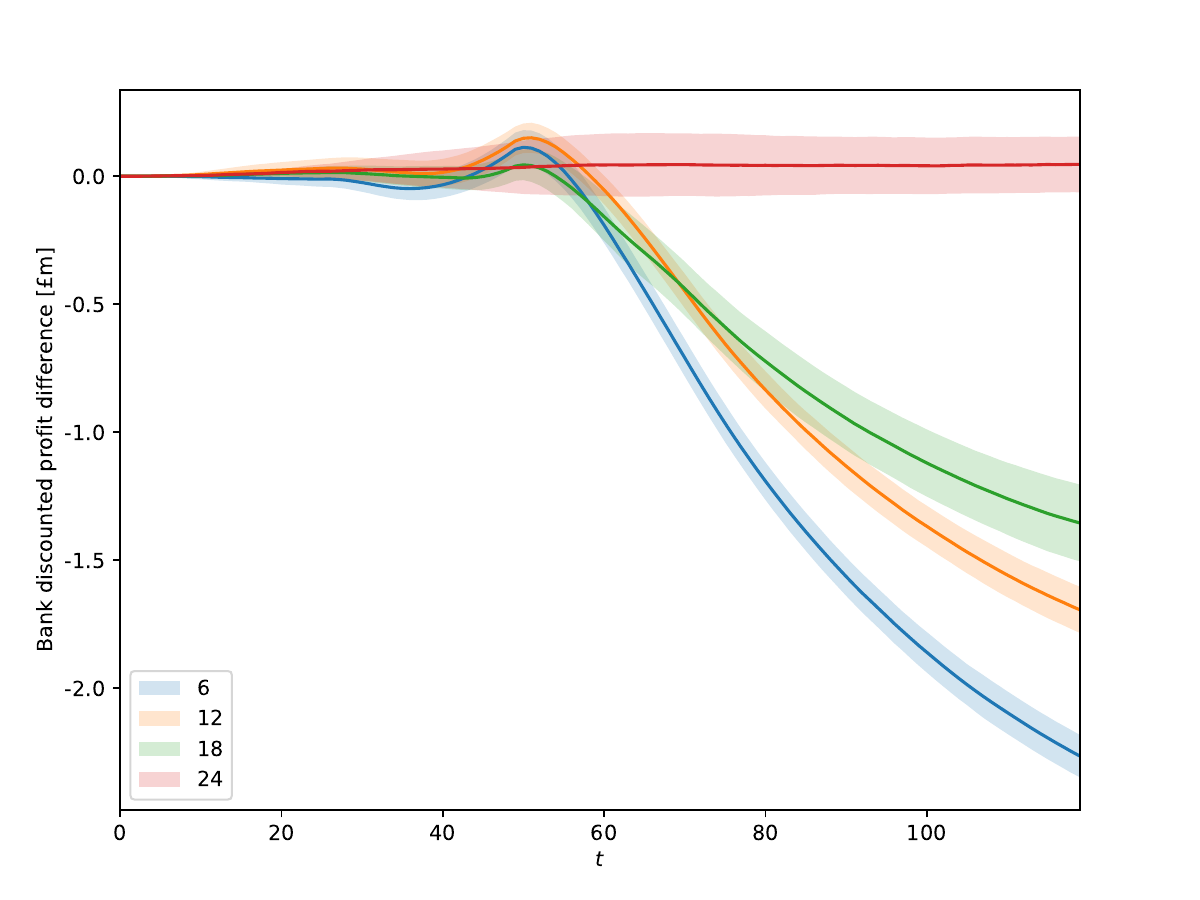}
		\caption{Difference in discounted profit of primary lender compared to a competitor offering a 24-month promotion.}
		\label{fig:Research_question_3_competitor_24_months_profit}
     \end{subfigure}
     \caption{Experimental results for a competing lender offering a 6-, 12-, 18-, 24- month promotion and the primary lender offering varying interest-free durations.
     Solid lines represent the mean and shaded areas correspond to one standard deviation.}
     \label{fig:Research_question_3_profit}
\end{figure}

\FloatBarrier

\begin{table}[ht]
\centering
\begin{threeparttable}
\caption{Percentage difference between discounted profit for primary and competing lenders for different interest-free durations.
Values are reported as the mean across five seeds, with uncertainties given as one standard deviation.
Bolded values are those with the largest mean percentage difference for each competing interest-free duration.}
\label{tab:RQ3_profit_percentage_difference_table}
\centering
\begin{tabular}{crrrr}
\hline
\diagbox[width=\dimexpr \textwidth/30+30\tabcolsep\relax]{Competing lender (months) }{Primary lender (months)} & 6 & 12 & 18 & 24  \\
\hline
6	&  0.1  $\pm$ 1.6	&  13.5 $\pm$  2.4	 & 15.1 $\pm$ 1.0	& \textbf{15.3 $\pm$ 2.2}  \\
12	& -11.7 $\pm$ 1.4	& -0.1  $\pm$  2.3 	& 11.6 $\pm$ 1.0	& \textbf{11.8 $\pm$ 2.0}  \\
18	& -13.4 $\pm$ 1.6	& -9.2  $\pm$  2.0	 & 0.4  $\pm$ 0.9	& \textbf{8.9 $\pm$ 1.4}  \\
24	& -13.0 $\pm$ 1.5	& -9.9  $\pm$  1.9 	& -8.1 $\pm$ 0.5	& \textbf{0.3 $\pm$ 1.6}  \\ \hline
\end{tabular}
\end{threeparttable}
\end{table}

\FloatBarrier

\subsection{Impact of time taken to respond to a competitor's promotion}
\begin{figure}
     \centering
     \begin{subfigure}[t]{0.45\textwidth}
		\centering
		\includegraphics[width=1.0\textwidth]{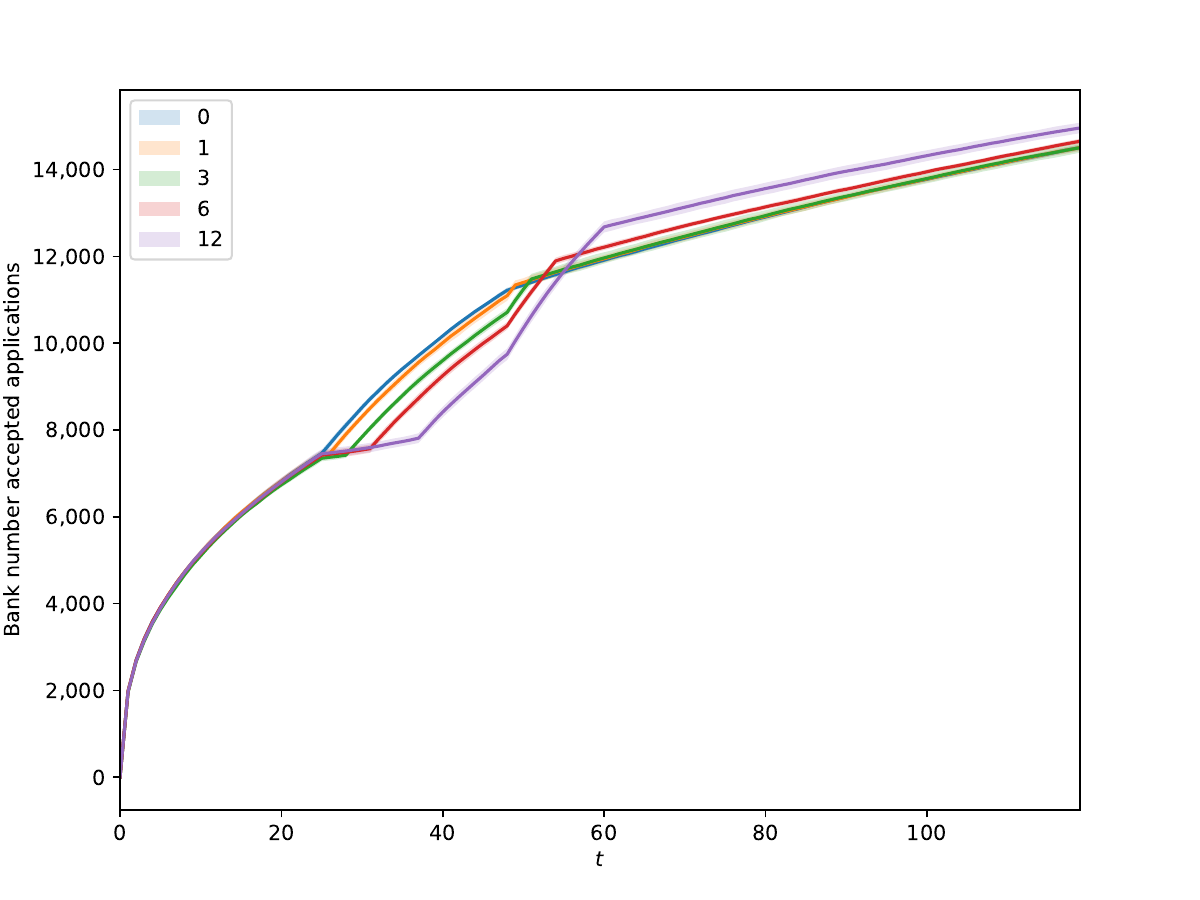}
		\caption{Cumulative number of accepted applications for the primary lender with varying promotion start times.}	
		\label{fig:Research_question_4_lender_number_applications}
     \end{subfigure}
     \hfill
     \begin{subfigure}[t]{0.45\textwidth}
		\centering		
		\includegraphics[width=1.0\textwidth]{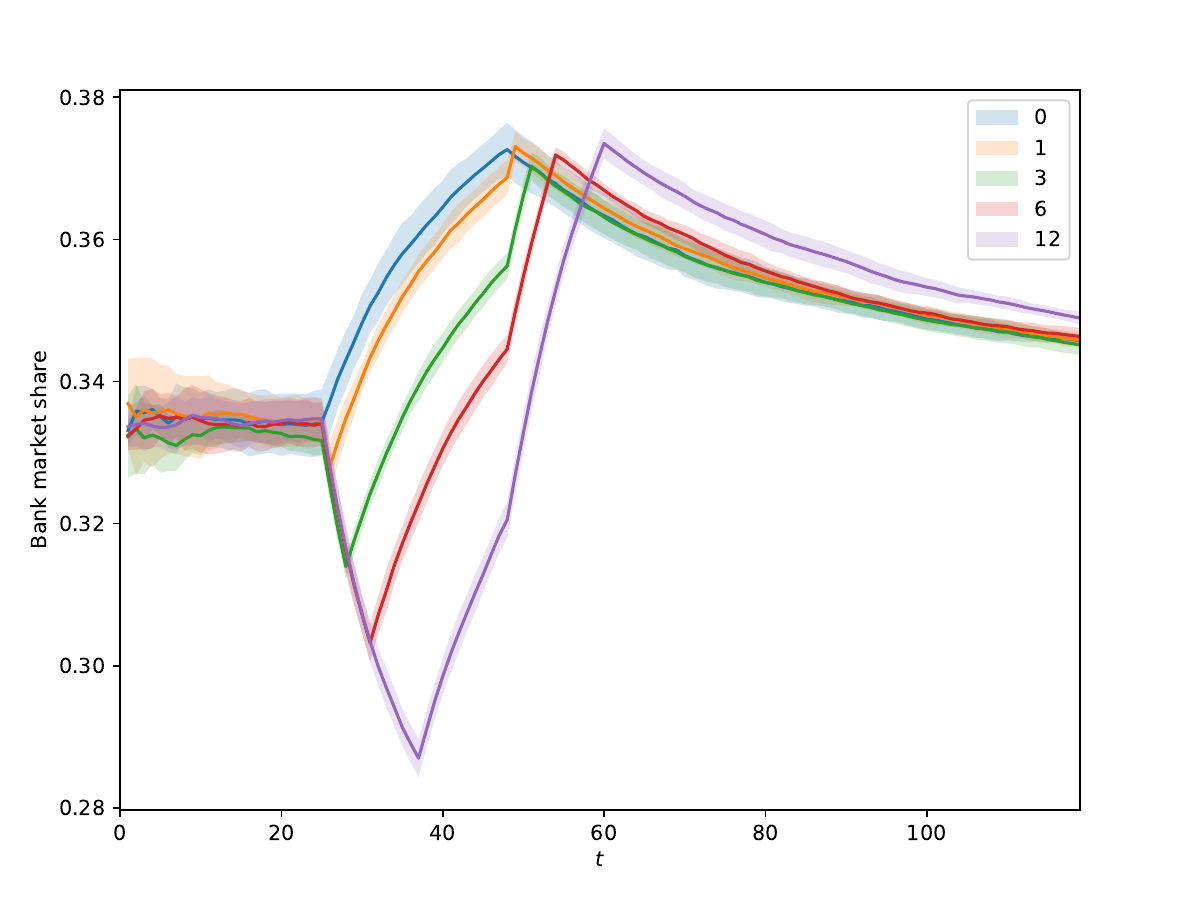}
		\caption{Market share for the primary lender with varying promotion start times.}
		\label{fig:Research_question_4_lender_market_share}
     \end{subfigure}
     \par\medskip
     \begin{subfigure}[t]{0.45\textwidth}
		\centering		
		\includegraphics[width=1.0\textwidth]{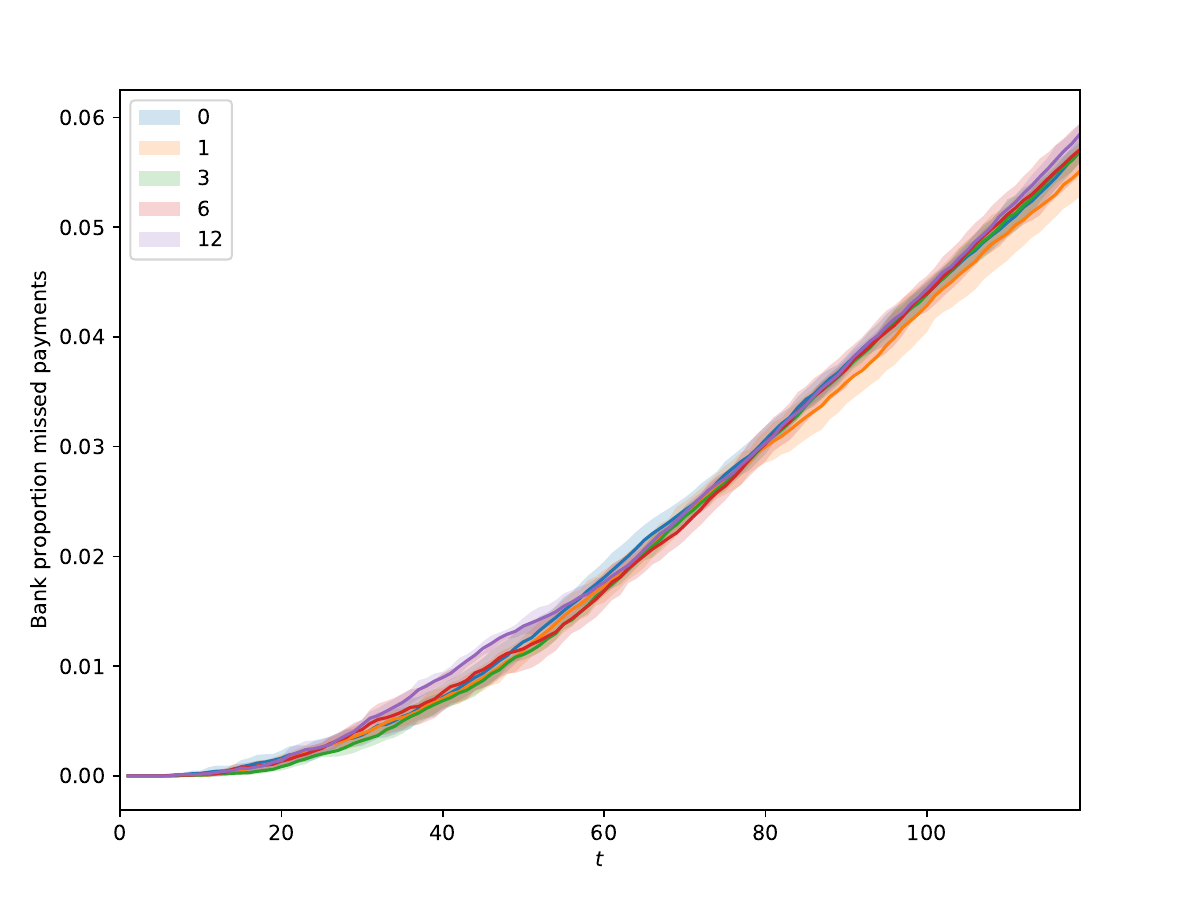}
		\caption{Proportion of cumulative missed payments for the primary lender with with varying promotion start times.}
		\label{fig:Research_question_4_missed_payments}
     \end{subfigure}
	 \hfill
     \begin{subfigure}[t]{0.45\textwidth}
		\centering		
		\includegraphics[width=1.0\textwidth]{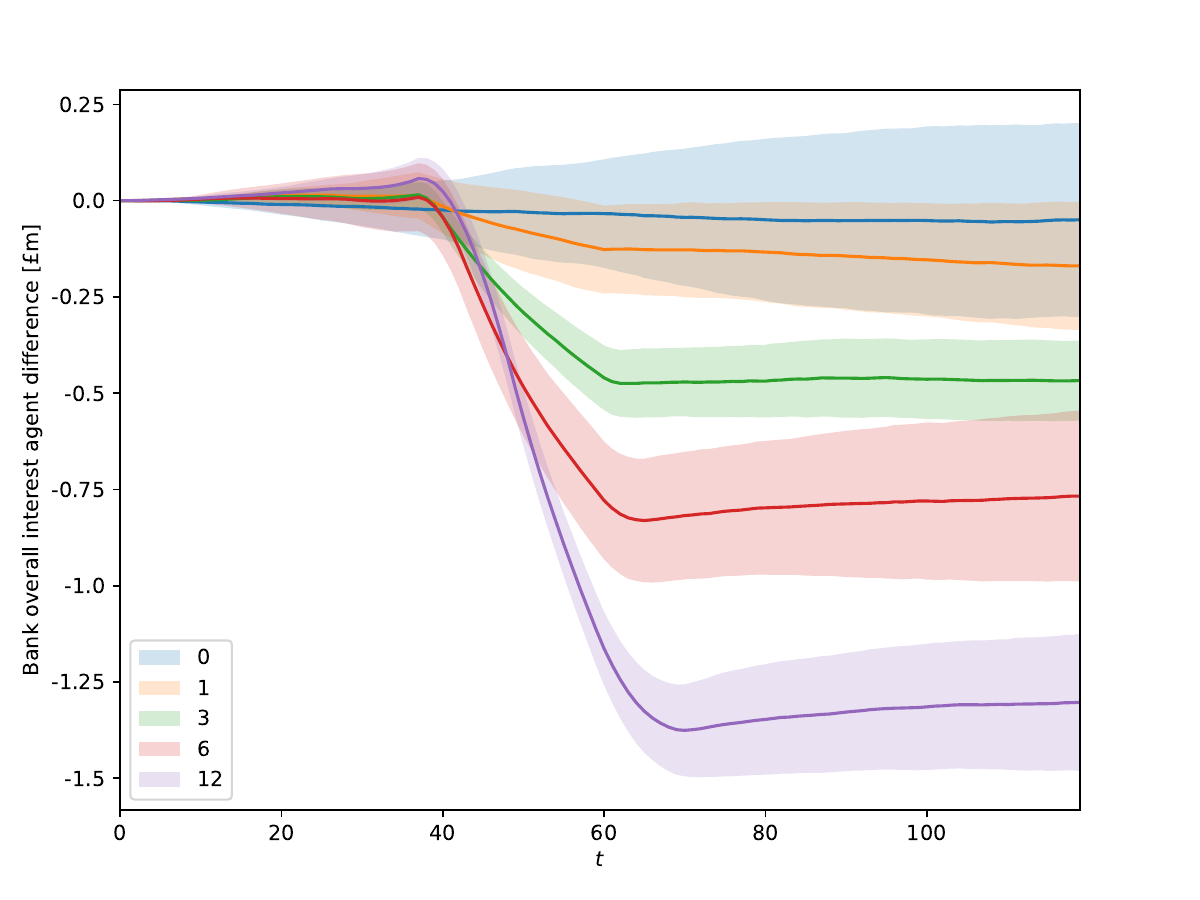}
		\caption{Cumulative discounted profit for the primary lender with varying promotion start times.}
		\label{fig:Research_question_4_lender_profit}
     \end{subfigure}
     \caption{Experimental results for the primary lender launching a promotional campaign at time steps of 24, 25, 27, 30, and 36 months, in reaction to a competitor's promotion launched at time $t$=24.
     The numbers in the legends show the time steps delay between the competing agent launching a promotion and the primary agent responding.
     Solid lines represent the mean and shaded areas correspond to one standard deviation.}
     \label{fig:Research_question_4}
\end{figure}

To examine the consequences of a delayed response to a competitor's promotional campaign, we conducted five distinct simulations. 
In each simulation, a competing lender initiated a promotion with a 12-month interest-free duration and a promotion-availability window of 24 months, commencing at $t$=24 and concluding at $t$=48. 
The primary lender, in response, launched a promotion with identical interest-free duration and promotion-availability window, albeit with a delay of 0, 1, 3, 6, and 12 months after the competitor's promotion initiation (corresponding to a promotion start at $t$=24, 25, 27, 30, 36). 
A third lender, offering no promotion, was also included in the simulations. 
All lenders applied a 20\% retail APR upon the conclusion of their respective promotions. 
The outcomes of these simulations are depicted in Figure \ref{fig:Research_question_4}.

As illustrated in Figure \ref{fig:Research_question_4_lender_number_applications}, the growth in successful applications for the primary lender decelerates whenever the competing lender is running a promotion and the primary lender is not. 
Short delays of 1 and 3 months do not significantly affect the trend of successful applications, but a delay of 12 months results in a more prolonged slowdown. 
Interestingly, towards the end of the simulation, the longest delay culminates in the highest number of successful applications, likely due to the period when the primary lender was the sole provider of a promotional offer.

The market share trends depicted in Figure \ref{fig:Research_question_4_lender_market_share} align with those of Figure \ref{fig:Research_question_4_lender_number_applications}. 
The later the primary lender's promotion commences, the more market share is lost, likely due to the competing lender with the earlier promotion securing a larger customer base. 
Upon the initiation of the primary lender's promotion, their market share increases and then gradually decreases towards the end of the simulation. 
Interestingly, the later the primary lender initiates their promotion, the larger their final market share, consistent with the later promotion start in Figure \ref{fig:Research_question_4_lender_number_applications} resulting in a larger cumulative number of accepted applications.
This can be attributed to the later start from the primary lender resulting in a longer period when they are the only lender offering a promotion, allowing them to recover market share.

Figure \ref{fig:Research_question_4_missed_payments} reveals that the longer the primary lender's response time, the greater the number of missed payments they are likely to encounter, with the slowest response periods resulting in the lowest proportion of missed payments. 
Figure \ref{fig:Research_question_4_lender_profit} compares the difference in profits between the primary and competitor lenders, indicating that whenever the delay in response to a promotion is 3 months or greater, the primary lender consistently records a lower profit than the competitor and fails to recover, even when the competitor has concluded their promotion.
This underscores that a slow response to a competitor's promotion can have long-term detrimental effects on a lender's profit, suggesting that lenders should promptly react to the launch of competing promotions. 
However, it should be noted that the magnitude of difference in profits is less than that shown in Figure \ref{fig:Research_question_3_competitor_24_months_profit}, suggesting that having a shorter interest-free duration than competitors can have a greater impact than a slow response.
The smaller magnitude of difference in profits also explains the relatively larger uncertainty bands observed in Figure \ref{fig:Research_question_4_lender_profit}.

This experiment investigates the implications of a lender's delayed response to a competing credit card promotion. 
A longer delay in launching a promotion in response to a competitor temporarily reduces market share, but over an extended period, it can result in a slightly larger market share (assuming the competing promotion will eventually end). 
However, a longer delay in responding to a competing promotion is also associated with an increase in the proportion of missed payments and a decrease in discounted profit, indicating that a delayed promotion response will increase risk and decrease profit for a lender. 
Overall, this experiment suggests that, for a lender primarily focused on long-term market share increase, a delayed response to a promotion is not of concern (assuming the competing promotion will end).
However, for any lender aiming to minimise their risk and maximise profits, minimising the delay in their response to a competing promotion is the optimal strategy.

\FloatBarrier

\section{Conclusion and Future Work}\label{sec:summary_label}

Promotional interest rates serve as a common strategy for attracting customers to new credit card products, thereby enhancing long-term lender profitability.
Despite the widespread use of these offers across various credit card markets, there is a noticeable gap in the literature regarding the analysis of these introductory promotions and their impact on customer behaviour.

The profitability derived from a single credit card product depends on a multitude of factors, including customer behaviour and the presence of competing offers from other credit card lenders.
Furthermore, promotional credit cards have several adjustable properties, making the analytical determination of an optimal credit card promotion a significant challenge.

In this paper, we have proposed and validated an agent-based model to examine the impact of different credit card promotions, demonstrating the feasibility of this method for exploring potential pricing strategies.
This model assigns agent attributes based on historical distributions and calibrates agent behaviours to align with historical benchmark values.
To the best of our knowledge, this is the first agent-based simulation focusing on credit card promotions with a strong emphasis on replicating and validating real-world trends and behaviours.

The model's ability to achieve a reasonable agreement with a variety of historical benchmarks concurrently (as demonstrated in Section \ref{sec:calibration_label}) underscores its ability to accurately replicate aspects of customer credit card usage.
We have validated this approach using two methods; replication of stylised facts and validation against time-series historical data from a UK credit card lender, Virgin Money.
The agreement with several stylised facts suggests that customer credit card behaviour is broadly and accurately replicated in our model.
Furthermore, there is agreement between the trends in historical time-wise data and the output from our simulation, including lender profit from interest, total customer balances, and lender market share.
This further validates the usefulness of our model for estimating the profit gained from a promotional credit card.

The experiments presented in this paper show how this model can be used to explore the effect of different aspects of a credit card promotion, namely the interest-free duration, the promotion-availability window, the interest-free duration offered in a competitive environment, and the time taken to respond to the launch of a competitor's promotion.
The effects of varying interest-free durations from a customer's perspective were also examined, including the impact of these promotions on the interest paid by customers, account balances, and the availability of convenient credit for customers.
The detailed analysis facilitated by agent-based modelling allows for the exploration of a variety of effects, including the impact of promotions on consumers, as well as the implications for revenue and capital requirements for lenders.

The robust foundation established by this paper paves the way for numerous research avenues to further enhance the model.
One such avenue involves the incorporation and refinement of mechanisms and attributes within the model.
For instance, customer agents could be assigned specific objectives, such as seeking short-term credit or aiming to improve their credit score, while lenders might be oriented towards maximising short-term profit or expanding their market share.
The model could also accommodate other credit card offers, such as balance transfer credit cards and cards offering cash back or loyalty points.
Additional revenue streams for providers, such as annual fees and interchange fees, could be integrated without necessitating a substantial overhaul of the model.
Moreover, enabling lenders to react to a competitor's promotion could facilitate the exploration of more intricate strategies.

Another promising direction for future research is the incorporation of more realistic price-response or demand behaviour. 
This could involve modelling how changes in the pricing of credit card promotions influence customer behaviour and demand, thereby providing a more nuanced understanding of the dynamics between pricing strategies and market responses.

Future work might also consider incorporating more stochastic agents and diverse behaviours.
As the irrational behaviour of credit card customers has been well documented in the literature \cite{Agarwal_2015}, including additional non-optimal decision making into lender actions could further improve the fidelity of the model.

The application of game theory for agent behaviour, as undertaken in the work of \citet{Andrade_2010}, could provide a promising direction for exploring lender pricing techniques in future research.
The integration of machine learning or reinforcement learning with agent-based modelling (ABM) approaches holds significant potential as an effective method for determining strategies.
Online reinforcement learning for optimising a pricing policy has proven effective in the agent-based approaches of \citet{Zhang_2007} and \citet{Han_2019}.
Recently, \citet{Khraishi_2022} demonstrated that offline reinforcement learning is an effective method for determining a pricing strategy for loans.
The application of one or more of these techniques could serve as a valuable extension to the model discussed in this work.
\subsubsection*{Acknowledgements}
We wish to express appreciation to Graham Smith and Zachery Anderson of NatWest Group for the time and support needed to develop this research paper. 

\bibliography{abm_library}
\appendix
\section{Appendix}
\label{sec:appendix}

\subsection{Customer attributes and behavioural parameters}\label{subsec:customer_attributes_label}
Every customer agent holds the following set of attributes:
	\paragraph{Age} -- sampled from the UK age distribution from 2020 ONS data.\footnote{\cite{ONS_Age_2020}} 
	Used to determine customer income and credit score, as described below.
	
	\paragraph{Income} $\income$ -- fixed amount per customer, representing their net annual income.
	A customer's gross income is sampled from 2020 ONS income data\footnote{\cite{ONS_Income_2020}}, using their age and a randomly sampled income percentile.
	The UK tax\footnote{\cite{Income_tax_2022}} and national insurance contributions\footnote{\cite{National_insurance_2022}} for the financial year ending in April 2020 are subtracted from the gross income to get customer net income.
	
	\paragraph{Total expenditure} $\total_exp$ -- fixed amount per customer, representing their annual expenditure. 
	Determined from the weekly expenditure data of one non-retired person household from 2020 ONS data.\footnote{\cite{ONS_Expenditure_2020}}
	For $\credexp$, all expenditure categories from the ONS data are considered.
	Appendix \ref{subsec:exp_categories} details the types of expenditure that are considered creditable and non-creditable.
	
	\paragraph{Creditable expenditure} $\credexp$ -- fixed amount per customer, representing the annual expenditure that can be paid for using a credit card. 
	The value is determined similarly to $\total_exp$, but only considering creditable expenditure categories, such as food, clothing and transport.
	
	\paragraph{Credit score} $\score$ -- a fixed value per customer, in $[300, 850]$ following the FICO score structure.\footnote{
	\href{https://www.fico.com/en/solutions/scoring-solutions}{https://www.fico.com}
	} $\score$ describes the customer's credit score at the start of the simulation and is used to represent historic credit card usage. 
	This value is determined using the probabilities of being in  different credit score ranges based on a customer's age; the probabilities are sourced from 2019 Statista data on the credit scores of 267 million U.S. Americans \citep{Statista_credit_score}.
	
	\paragraph{Credit cards owned} -- a set of credit card objects that the customer owns; a customer's total overall balance at any given time may be obtained as the sum of the credit cards' balances.
	The maximum number of cards is equal to the number of lenders in a simulation, which is 3 for the simulations described in this work.
	
	\paragraph{Comfort limit} -- a constant parameter across customers with values in $[0, 1]$, describing the proportion of the credit limit the customer prefers not to exceed in credit card balance. 
	This limit is exceeded by a customer only when necessary and is taken to be 30\% of the card's credit limit, as the maximum credit utilisation advised by credit bureaus.\footnote{
	\href{https://www.experian.com/blogs/ask-experian/credit-education/score-basics/credit-utilization-rate/}{https://www.experian.com}
	} 
		
	\paragraph{Application constant} $\rho$ -- a constant parameter across customers, which controls the relationship between a customer's likelihood to apply to another card; obtained through calibration.
	
	\paragraph{Better offer weighting} $\better_offer_weighting$ -- a constant parameter across customers, which controls the effect of a customer having better offers on their current credit card on the likelihood of them applying to a new one; obtained through calibration.
	
	\paragraph{Application tendency} $\apptend$ -- a constant parameter across customers, which describes an independent psychological factor which controls the relationship between a customer’s likelihood to apply and the difference between their expenditure and income; obtained through calibration.
	
	\paragraph{Application minimum probability} $\appminprob$ -- a constant parameter across customers, which describes the minimum probability each step that a customer will apply to a credit card, regardless of the difference between their income and expenditure; obtained through calibration.
	
	\paragraph{Application rationality} $\apprationality$ -- a constant parameter across customers, which controls how likely a customer is to make the rational (i.e. cheapest) decision when selecting a new credit card; obtained through calibration.
	
	\paragraph{Application limit weighting} $\applimitweighting$ -- a constant parameter across customers, which controls the relationship between how likely a customer is to apply for a new card and how close on average they are to the credit limits of the cards they own; obtained through calibration.
	
	\paragraph{Repayment tendency} $\repaytend$ -- a constant parameter across customers, which describes the relationship between a customer's credit score and how much of their credit card balance customers will repay; obtained through calibration.
	
	\paragraph{Repayment constant} $\repayconst$ -- a constant parameter across customers, which controls how much a customer will pay off their credit cards each turn; obtained through calibration.
	
	\paragraph{Credit limit weighting}  $\cred_lim_weighting$ -- a constant parameter across customers, which controls the effect of how close a customer is to their credit limits and how much money they spend on their credit cards; obtained through calibration.
	
	\paragraph{Probability of missing minimum payment} $\miss_pay_prob$ -- a constant parameter across customers, the likelihood of a customer missing a minimum payment on a credit card; obtained through calibration.
		
	\paragraph{Usage tendency} $\usage_tendency$ -- a constant parameter across customers, which describes an independent psychological factor giving the likelihood of credit card usage amongst the expenditure that is available for credit card usage; a customer's usage tendency parameter is obtained through calibration.
	
	\paragraph{Interest rate payment weighting} $\interestweighting$ -- a constant parameter across customers, which controls how strongly a customer prioritises interest rate when choosing what card to use; obtained through calibration.
	
	\paragraph{Balance payment weighting} $\balanceweighting$ -- a constant parameter across customers, which controls how strongly a customer prioritises balance when choosing what card to use; obtained through calibration.

\subsubsection{Expenditure categories}\label{subsec:exp_categories}

Creditable and total expenditure categories used in this model have been determined based on ONS data of household spending in the UK \cite{ONS_Expenditure_2020}. 

Creditable expenditure ($\credexp$) has been defined using the following payment categories:
\begin{itemize}
	\item Food and non-alcoholic drinks
	\item Alcoholic drink, tobacco and narcotics
	\item Clothing and footwear
	\item Health
	\item Transport
	\item Communication
	\item Recreation and culture
	\item Education
	\item Restaurants and hotels
	\item Miscellaneous goods and services
\end{itemize}

Total expenditure ($\total_exp$) includes the creditable expenditure categories detailed above, in addition to the following non-creditable expenditure categories:
\begin{itemize}
	\item Housing (net), fuel and power
	\item Household goods and services
	\item Other expenditure items
\end{itemize}

\FloatBarrier
\subsubsection{Histograms of customer agent attributes}
\FloatBarrier
\begin{figure}[ht!]
     \centering
     \begin{subfigure}[t]{0.45\textwidth}
	 	\centering
		\includegraphics[width=1.0\textwidth]{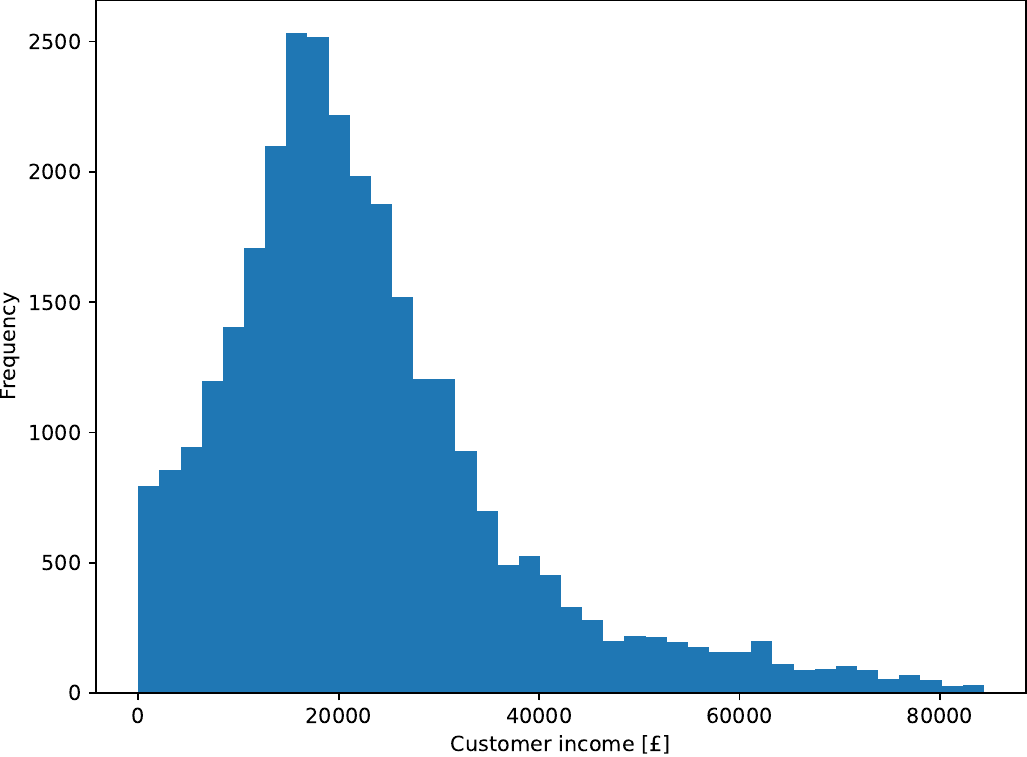} 
		\caption{Histogram of (net) annual income of customer agents.}
     \end{subfigure}
     \hfill
     \begin{subfigure}[t]{0.45\textwidth}
     	\centering
		\includegraphics[width=1.0\textwidth]{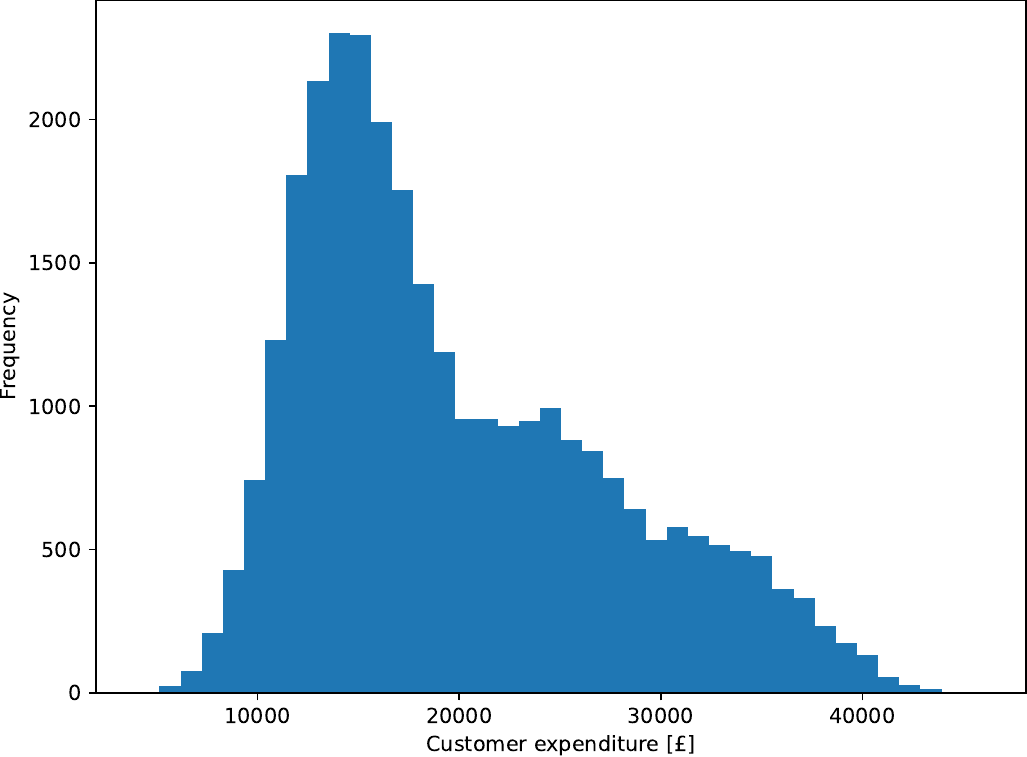}
		\caption{Histogram of annual total expenditure of customer agents.}
     \end{subfigure}
     \par\medskip
     \begin{subfigure}[t]{0.45\textwidth}
     	\centering
		\includegraphics[width=1.0\textwidth]{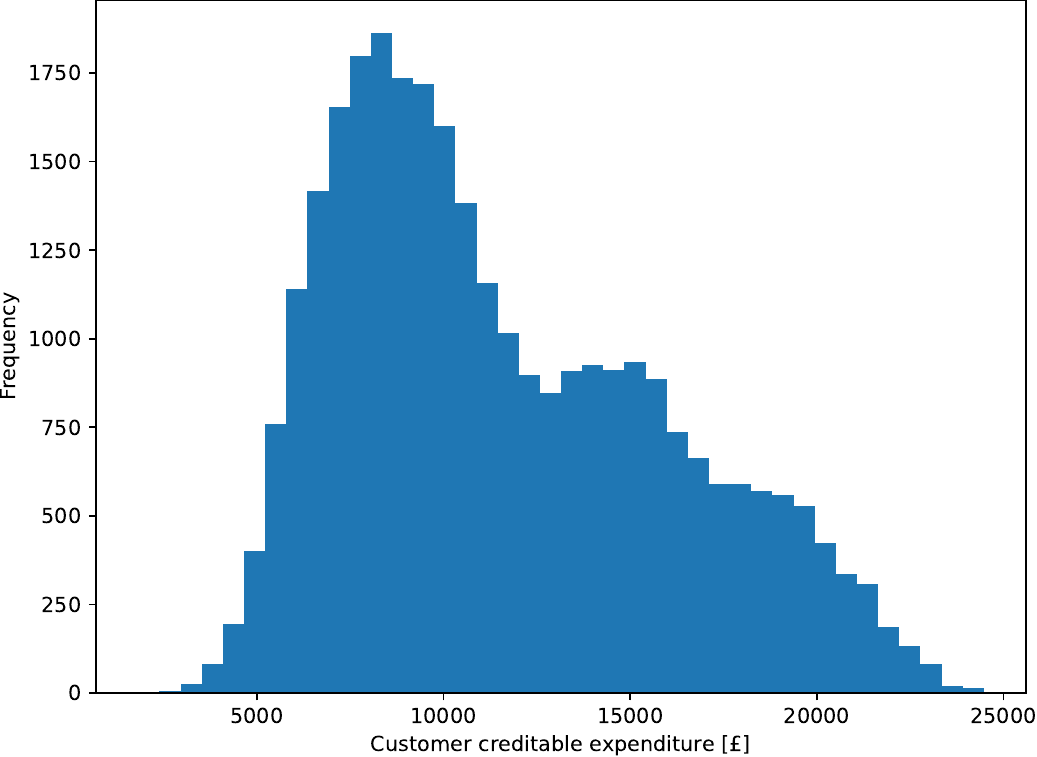}
		\caption{Histogram of creditable expenditure of customer agents.}
     \end{subfigure}
	 \hfill     
     \begin{subfigure}[t]{0.45\textwidth}
     	\centering
		\includegraphics[width=1.0\textwidth]{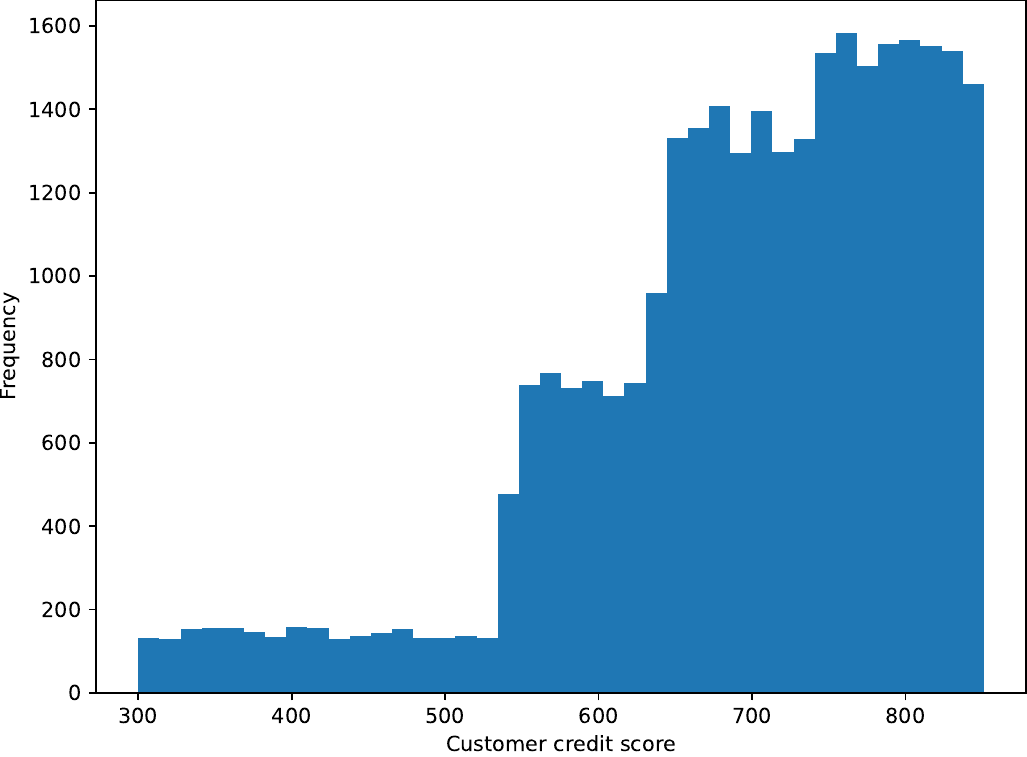}
		\caption{Histogram of credit score of customer agents.}
     \end{subfigure}
     \caption{Histograms of attributes of 30,000 customer agents.}
     \label{fig:customer_attribute_hists}
\end{figure}
	
\FloatBarrier

\subsubsection{Customer card repayment algorithms}\label{subsec:repayment_algo}

\begin{algorithm}
\caption{Repayment of minimum payments on cards.}
\label{alg:min_payments}
\begin{algorithmic}[1]
\Statex \textbf{Input:} cards owned by customer $j$ $n_j$, probability of missing minimum payment $P_{\text{miss}}$.
\Procedure{RepayMinimumPayments}{$n_j$; $P_{\text{miss}}$}
	\For {$k=1; k \le n_j; k++$}
		\State $\xi \sim U(0, 1)$
		\If{$\xi > P_{\text{miss}}$}
			\State $\balance_{j, k} \gets \balance_{j, k} - M_{j, k}$
			\State $BUDGET_j \gets BUDGET_j - M_{j, k}$
		\EndIf
	\EndFor
\EndProcedure
\end{algorithmic}
\end{algorithm}

\begin{algorithm}
\caption{Repayment of cards after minimum payments have been made.}
\label{alg:repayment}
\begin{algorithmic}[1]
\Statex \textbf{Input:} cards owned by customer $j$ $n_j$, budget for repayment from customer $j$ $BUDGET_{j}$.
\Procedure{RepayBalances}{$n_j$; $BUDGET_{j}$}
	\For {$k=1; k \le n_j; k++$}
		\State $PAY = \text{min}(BUDGET_j, \balance_{j, k})$
		\State $\balance_{j, k} \gets \balance_{j, k}- PAY$
		\State $BUDGET_j \gets BUDGET_j - PAY$
	\EndFor
\EndProcedure
\end{algorithmic}
\end{algorithm}

\FloatBarrier

\subsection{Lender attributes and behavioural parameters}
Each credit card lender has the following attributes:
	\paragraph{Acceptance Criteria} -- the acceptance criteria that a lender has for accepting a customer's credit card application; this involves a combination of thresholds for customer credit score and customer income which are set before each simulation.
	
	\paragraph{Credit Limit} $\credlimit$ -- the credit limit a lender can offer based on the customer's income. 
	The balance of the card cannot exceed this value and is set whenever the card is taken out by a customer.

	\paragraph{Promotion Strategy} -- the promotion strategy defined in terms of interest-free period duration, promotion-availability window of the campaign, and the time step that the promotion begins at; this is set before each simulation.

	\paragraph{Credit Card Offered} -- a credit card object whose attributes are determined by the promotion strategy and the applicant's attributes.

	\paragraph{Profit} $PR_t$ -- sum of the cumulative interest and fees charged up to time $t$ across credit cards associated with the lender.
	Interest profit of a lender agent at time step $t$ is the cumulative sum of the interest charged to accounts up to time step $t$.
	In our model, non-interest profit of a lender at time step $t$ is the sum of late payment fees charged to customers.

	\paragraph{Discounted profit} $DPR_t$ -- sum of discounted profits up to a time $t$ for a lender.
	Discounted profit is determined as 
\begin{equation}
	DPR_t = \sum_{t=0}^{T} \frac{PR_{t}}{(1 + r)^{t}},
\end{equation}	
	 where $PR_{t}$ is the profit gained at time step $t$, and $r$ is the discount rate, set at 3.5\% APR per year (12 time steps) as recommended by \citet{Freeman_2017}.

\subsection{Experimental results}\label{subsec:experiment_results}

\FloatBarrier

\subsubsection{Model stability experiments}
\begin{figure}
	\centering
	\includegraphics[scale=0.6]{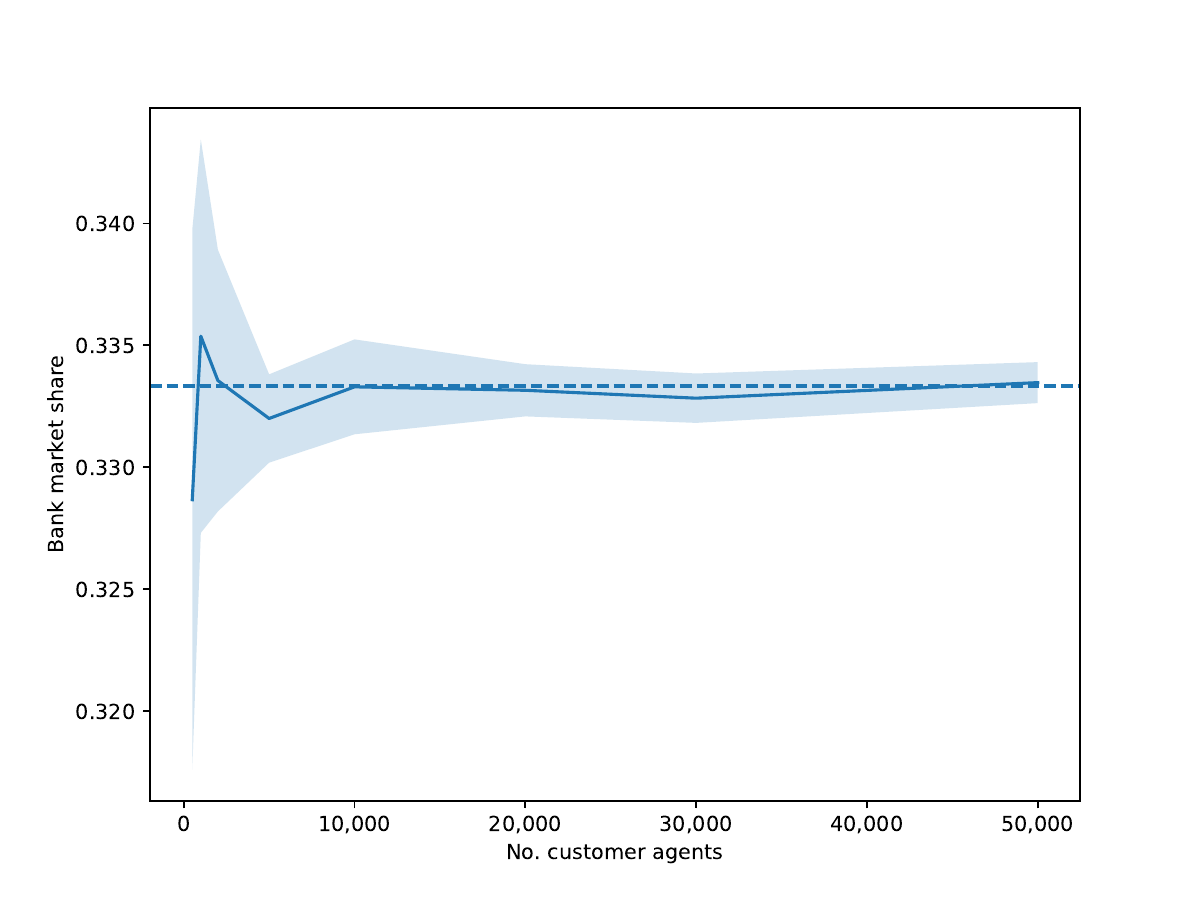}
	\caption{Market share for a lender agent across a range of numbers of customer agents. 
	Simulations were ran across five seeds, with the solid lines representing the mean of the the market share and the shaded areas representing one standard deviation across the seeds.
	The dashed horizontal line indicates a market share of a third, the expected value for three identical lenders.}
	\label{figure:stability_experiment_plot}
\end{figure}

To estimate how many customer agents are required for a stable model output, simulations were ran with 3 lender agents, all with no promotion.
For this setup, as the 3 lenders are identical, a market share of a third each is expected.
Simulations were ran for numbers of customer agents in the range 500-50,000, across 5 random seeds.
The market share of a lender in these simulations is shown in Figure \ref{figure:stability_experiment_plot} as a function of the number of customer agents.

\begin{figure}
	\centering
	\includegraphics[scale=0.6]{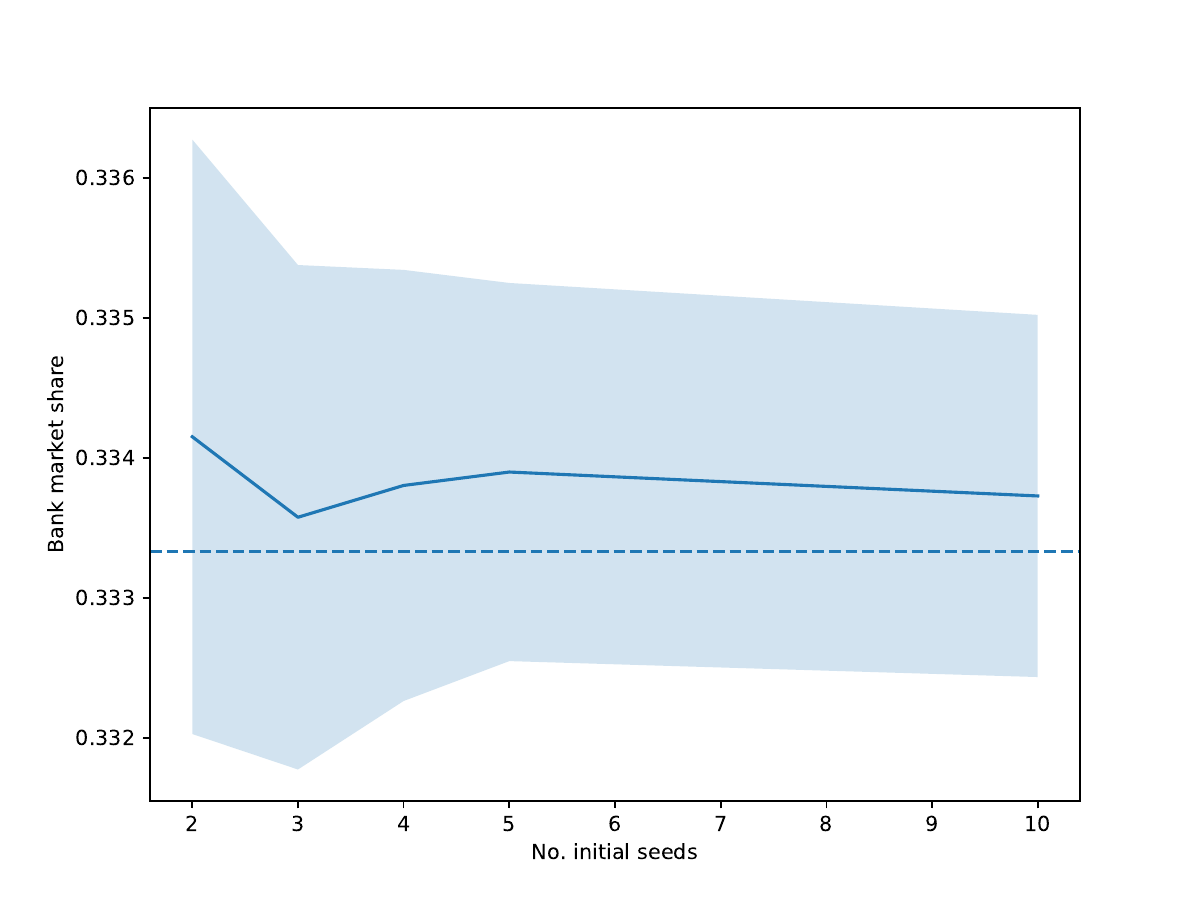}
	\caption{Market share for a lender agent across a range of numbers of initial seeds. 
	Simulations were ran using 30,000 agents, with the solid lines representing the mean of the the market share and the shaded areas representing one standard deviation across the seeds.
	The dashed horizontal line indicates a market share of a third, the expected value for three identical lenders.}
	\label{figure:num_seeds_stability_experiment_plot}
\end{figure}

Additionally, an experiment was ran exploring the stability of the model with different numbers of initial seeds.
Similar to the experiment described above, models with 3 identical lenders were ran, with a market share of a third expected for each lender.
Simulations were ran for numbers of initial random seeds in the range 2-10, with 30,000 agents. 
The market share of a lender in these simulations is shown in Figure \ref{figure:num_seeds_stability_experiment_plot} as a function of the number of initial seeds.

\FloatBarrier

\subsubsection{Effect of credit card promotion interest-free duration on optimising a lender's pricing strategy}

\begin{table}[ht]
\centering
\begin{threeparttable}
\caption{Experimental results for lenders with different interest-free durations. Differences are represented as the mean percentage difference relative to the competitor lender agent, with uncertainties given as the standard deviation across five seeds.}
\label{tab:RQ1-table}
\centering
\begin{tabular}{llrrr}
\toprule
Simulation & \begin{tabular}[l]{@{}l@{}}Lender interest-free\\ duration\end{tabular} & \begin{tabular}[r]{@{}r@{}}Lender discounted\\ interest difference\end{tabular} & \begin{tabular}[r]{@{}r@{}}Lender fraction missed\\ payments difference\end{tabular} & \begin{tabular}[r]{@{}r@{}}Lender\\ market share\end{tabular} \\
\midrule
           &                    None &                          -0.1 ± 1.9 &                                      1 ± 8 &   0.3327 ± 0.0022 \\
         1 &                    None &                             0.0 ± 0.8 &                                      0 ± 8 &   0.3336 ± 0.0014 \\
           &                    None &                             0.0 ± 1.2 &                                      0 ± 7 &   0.3337 ± 0.0013 \\
\midrule
           &                    None &                           0.4 ± 2.4 &                                 -0.2 ± 2.7 &   0.3174 ± 0.0010 \\
         2 &                    None &                             0.0 ± 2.5 &                                      0 ± 3 &   0.3173 ± 0.0011 \\
           &                       6 &                              18 ± 3 &                                -10.8 ± 2.3 &   0.3653 ± 0.0010 \\
\midrule
           &                    None &                           0.2 ± 1.3 &                                    0.0 ± 2.9 &   0.3118 ± 0.0005 \\
         3 &                    None &                             0.0 ± 1.6 &                                      0 ± 4 &   0.3111 ± 0.0006 \\
           &                      12 &                          19.8 ± 1.8 &                                -14.6 ± 2.8 &   0.3771 ± 0.0007 \\
\midrule
           &                    None &                             0.1 ± 2.0 &                                    0 ± 3 &   0.2979 ± 0.0007 \\
         4 &                    None &                             0.0 ± 1.8 &                                      0 ± 3 &   0.2984 ± 0.0009 \\
           &                      24 &                          19.4 ± 1.1 &                                -22.1 ± 1.9 &   0.4037 ± 0.0006 \\
\midrule
           &                    None &                             0.0 ± 1.5 &                                     -1 ± 3 &   0.2889 ± 0.0009 \\
         5 &                    None &                             0.0 ± 1.9 &                                      0 ± 3 &    0.2880 ± 0.0018 \\
           &                      36 &                          13.6 ± 1.5 &                                -27.8 ± 2.1 &    0.4230 ± 0.0010 \\
\midrule
           &                    None &                           0.1 ± 1.6 &                                   -0.4 ± 1.0 &   0.2855 ± 0.0007 \\
         6 &                    None &                             0.0 ± 1.5 &                                    0.0 ± 0.9 &    0.2860 ± 0.0007 \\
           &                      48 &                           4.9 ± 1.4 &                                  -29.9 ± 1.0 &   0.4286 ± 0.0010 \\
\midrule
\bottomrule
\end{tabular}
\end{threeparttable}
\end{table}

\FloatBarrier

\newpage
\subsubsection{Effect of credit card promotion-availability windows on optimising a lender's pricing strategy}

\begin{table}[ht]
\centering
\begin{threeparttable}
\caption{Experimental results for lenders with different promotion-availability windows. 
Differences are represented as the mean percentage difference relative to the competitor lender agent, with uncertainties given as the  standard deviation across five seeds.}
\label{tab:RQ2-table}
\centering
\begin{tabular}{llrrr}
\toprule
Simulation & \begin{tabular}[l]{@{}l@{}}Lender promotion\\ availability window \end{tabular} & \begin{tabular}[r]{@{}r@{}}Lender discounted\\ interest difference\end{tabular} & \begin{tabular}[r]{@{}r@{}}Lender fraction missed\\ payments difference\end{tabular} & \begin{tabular}[r]{@{}r@{}}Lender\\ market share\end{tabular} \\
\midrule
           &                 None &                          -0.1 ± 1.9 &                                      1 ± 8 &   0.3327 ± 0.0022 \\
         1 &                 None &                             0.0 ± 0.8 &                                      0 ± 8 &   0.3336 ± 0.0014 \\
           &                 None &                             0.0 ± 1.2 &                                      0 ± 7 &   0.3337 ± 0.0013 \\
\midrule
           &                 None &                             0.0 ± 2.7 &                                      0 ± 3 &   0.3272 ± 0.0007 \\
         2 &                 None &                             0.0 ± 2.4 &                                      0 ± 4 &   0.3281 ± 0.0009 \\
           &                    6 &                           6.9 ± 2.8 &                                   -4 ± 3 &   0.3447 ± 0.0013 \\
\midrule
           &                 None &                           0.2 ± 1.5 &                                    0.0 ± 2.5 &   0.3227 ± 0.0005 \\
         3 &                 None &                             0.0 ± 1.2 &                                    0.0 ± 2.6 &   0.3214 ± 0.0005 \\
           &                   12 &                          11.2 ± 1.7 &                                 -7.6 ± 2.5 &    0.3560 ± 0.0004 \\
\midrule
           &                 None &                           0.0 ± 1.8 &                                      0 ± 5 &   0.3116 ± 0.0015 \\
         4 &                 None &                             0.0 ± 1.6 &                                      0 ± 4 &   0.3118 ± 0.0012 \\
           &                   24 &                          19.8 ± 0.9 &                                    -14 ± 4 &   0.3766 ± 0.0014 \\
\midrule
           &                 None &                          -0.3 ± 1.3 &                                 -0.4 ± 2.2 &   0.3029 ± 0.0014 \\
         5 &                 None &                             0.0 ± 1.7 &                                    0.0 ± 2.5 &   0.3035 ± 0.0004 \\
           &                   36 &                          25.5 ± 1.8 &                                -19.6 ± 2.7 &   0.3936 ± 0.0019 \\
\midrule
           &                 None &                           0.2 ± 1.8 &                                      1 ± 4 &   0.2953 ± 0.0007 \\
         6 &                 None &                             0.0 ± 1.7 &                                      0 ± 4 &   0.2962 ± 0.0010 \\
           &                   48 &                          29.2 ± 1.7 &                                -23.2 ± 2.7 &   0.4085 ± 0.0009 \\
\midrule
\bottomrule
\end{tabular}
\end{threeparttable}
\end{table}

\FloatBarrier
\subsubsection{Effect of competing credit card promotions on the optimal strategy for a lender with a promotion}

\begin{table}[ht]
\small
\centering
\begin{threeparttable}
\caption{Experimental results for different primary lender and competing lender interest-free durations. 
Differences are represented as the mean percentage difference relative to the competitor lender agent, with uncertainties given as the  standard deviation across five seeds.}
\label{tab:RQ3-table}
\centering
\begin{tabular}{llrrr}
\toprule
Simulation & \begin{tabular}[l]{@{}l@{}}Lender interest-free\\ duration\end{tabular} & \begin{tabular}[r]{@{}r@{}}Lender discounted\\ interest difference\end{tabular} & \begin{tabular}[r]{@{}r@{}}Lender fraction missed\\ payments difference\end{tabular} & \begin{tabular}[r]{@{}r@{}}Lender\\ market share\end{tabular} \\
\midrule
           &                    None &                         -11.3 ± 1.5 &                                      7 ± 6 &   0.3131 ± 0.0017 \\
         1 &                       6 &                             0.0 ± 1.6 &                                      0.0 ± 4 &   0.3434 ± 0.0011 \\
           &                       6 &                           0.1 ± 1.6 &                                    0.0 ± 4 &   0.3435 ± 0.0015 \\
\midrule
           &                    None &                          -5.7 ± 2.7 &                                    3.9 ± 2.0 &   0.3079 ± 0.0010 \\
         2 &                       6 &                             0.0 ± 2.4 &                                      0 ± 2.0 &   0.3217 ± 0.0006 \\
           &                      12 &                          13.5 ± 2.4 &                                -10.6 ± 0.9 &   0.3704 ± 0.0006 \\
\midrule
           &                    None &                          -5.2 ± 1.4 &                                    2.6 ± 3 &   0.3036 ± 0.0015 \\
         3 &                       6 &                             0.0 ± 1.2 &                                      0 ± 3 &   0.3131 ± 0.0011 \\
           &                      18 &                            15.1 ± 1.0 &                                -15.6 ± 2.7 &   0.3833 ± 0.0020 \\
\midrule
           &                    None &                          -3.7 ± 1.7 &                                    1.8 ± 2.0 &   0.2979 ± 0.0005 \\
         4 &                       6 &                             0.0 ± 1.9 &                                    0.0 ± 1.2 &    0.3060 ± 0.0010 \\
           &                      24 &                          15.2 ± 2.2 &                                -19.6 ± 1.4 &    0.3960 ± 0.0011 \\
\midrule
           &                    None &                         -16.7 ± 1.7 &                                     15 ± 6 &   0.3099 ± 0.0025 \\
         5 &                      12 &                             0.0 ± 1.6 &                                      0 ± 4 &   0.3695 ± 0.0016 \\
           &                       6 &                         -11.7 ± 1.4 &                                     12 ± 5 &   0.3206 ± 0.0022 \\
\midrule
           &                    None &                         -11.7 ± 2.2 &                                     10 ± 4 &   0.3087 ± 0.0025 \\
         6 &                      12 &                             0.0 ± 2.7 &                                      0 ± 5 &   0.3457 ± 0.0017 \\
           &                      12 &                          -0.1 ± 2.3 &                                      0 ± 4 &   0.3456 ± 0.0010 \\
\midrule
           &                    None &                          -7.1 ± 1.7 &                                      4 ± 4 &   0.3035 ± 0.0017 \\
         7 &                      12 &                             0.0 ± 1.2 &                                      0 ± 4 &  0.3201 ± 0.0017 \\
           &                      18 &                            11.6 ± 1.0 &                                    -13 ± 3 &   0.3764 ± 0.0007 \\
\midrule
           &                    None &                          -5.9 ± 1.9 &                                      5 ± 6 &   0.2974 ± 0.0010 \\
         8 &                      12 &                             0.0 ± 1.8 &                                      0 ± 5 &   0.3135 ± 0.0015 \\
           &                      24 &                            11.8 ± 2.0 &                                    -16 ± 4 &   0.3891 ± 0.0017 \\
\midrule
           &                    None &                         -17.5 ± 1.4 &                                     22 ± 4 &   0.3032 ± 0.0010 \\
         9 &                      18 &                             0.0 ± 1.6 &                                      0 ± 4 &   0.3832 ± 0.0005 \\
           &                       6 &                         -13.4 ± 1.6 &                                     17 ± 4 &   0.3136 ± 0.0012 \\
\midrule
           &                    None &                         -15.6 ± 1.9 &                                 19.6 ± 1.9 &   0.3041 ± 0.0013 \\
        10 &                      18 &                             0.0 ± 2.2 &                                    0.0 ± 1.1 &   0.3761 ± 0.0013 \\
           &                      12 &                          -9.2 ± 1.9 &                                 14.1 ± 1.8 &   0.3198 ± 0.0012 \\
\midrule
           &                    None &                         -11.8 ± 1.6 &                                 12.4 ± 2.9 &   0.3038 ± 0.0018 \\
        11 &                      18 &                             0.0 ± 0.8 &                                    0.0 ± 2.7 &   0.3479 ± 0.0012 \\
           &                      18 &                           0.4 ± 0.9 &                                  0.3 ± 2.5 &   0.3483 ± 0.0011 \\
\midrule
           &                    None &                          -7.3 ± 1.7 &                                      6 ± 5 &   0.2982 ± 0.0014 \\
        12 &                      18 &                             0.0 ± 1.5 &                                      0 ± 4 &   0.3208 ± 0.0010 \\
           &                      24 &                           8.9 ± 1.4 &                                    -13 ± 3 &    0.3810 ± 0.0016 \\
\midrule
           &                    None &                         -16.2 ± 1.4 &                                     27 ± 5 &   0.2973 ± 0.0015 \\
        13 &                      24 &                             0.0 ± 1.5 &                                      0 ± 4 &   0.3971 ± 0.0017 \\
           &                       6 &                           -13 ± 1.5 &                                     24 ± 5 &   0.3056 ± 0.0008 \\
\midrule
           &                    None &                         -15.2 ± 1.9 &                                 24.7 ± 2.8 &   0.2981 ± 0.0006 \\
        14 &                      24 &                             0.0 ± 2.1 &                                    0.0 ± 1.8 &   0.3895 ± 0.0011 \\
           &                      12 &                          -9.9 ± 1.9 &                                 20.1 ± 2.6 &   0.3123 ± 0.0010 \\
\midrule
           &                    None &                           -15.0 ± 1.3 &                                 22.7 ± 2.5 &   0.2979 ± 0.0011 \\
        15 &                      24 &                             0.0 ± 0.9 &                                    0.0 ± 1.6 &   0.3814 ± 0.0018 \\
           &                      18 &                          -8.1 ± 0.5 &                                 15.9 ± 1.7 &   0.3207 ± 0.0017 \\
\midrule
           &                    None &                         -10.6 ± 1.7 &                                   15 ± 3 &   0.2972 ± 0.0005 \\
        16 &                      24 &                             0.0 ± 1.5 &                                    0.0 ± 2.5 &    0.3510 ± 0.0007 \\
           &                      24 &                           0.3 ± 1.6 &                                 -0.6 ± 2.3 &   0.3518 ± 0.0006 \\
\midrule
\bottomrule
\end{tabular}
\end{threeparttable}
\end{table}

\FloatBarrier

\subsubsection{Effect of time taken to respond to a competitor's promotion on a promotion campaign}

\begin{table}[ht]
\centering
\begin{threeparttable}
\caption{Experimental results for different lender reaction times. 
Differences are represented as the mean percentage difference relative to the competitor lender agent, with uncertainties given as the  standard deviation across five seeds.}
\label{tab:RQ4-table}
\centering
\begin{tabular}{llrrr}
\toprule
Simulation & \begin{tabular}[l]{@{}l@{}}Lender promotion\\ start month\end{tabular} & \begin{tabular}[r]{@{}r@{}}Lender discounted\\ interest difference\end{tabular} & \begin{tabular}[r]{@{}r@{}}Lender fraction missed\\ payments difference\end{tabular} & \begin{tabular}[r]{@{}r@{}}Lender\\ market share\end{tabular} \\
\midrule
           &                 None &                         -12.2 ± 2.1 &                                  9.1 ± 2.6 &   0.3095 ± 0.0010 \\
         1 &                   24 &                           0.0 ± 2.7 &                                    0.0 ± 2.4 &   0.3447 ± 0.0013 \\
           &                   24 &                          -0.3 ± 1.5 &                                    0.0 ± 2.1 &   0.3458 ± 0.0006 \\
\midrule
           &                 None &                           -13.0 ± 1.9 &                                      8 ± 4 &    0.3090 ± 0.0014 \\
         2 &                   24 &                           0.0 ± 1.9 &                                      0.0 ± 4 &   0.3451 ± 0.0009 \\
           &                   25 &                          -0.9 ± 1.9 &                                    0.0 ± 4 &   0.3459 ± 0.0011 \\
\midrule
           &                 None &                           -13.9 ± 1.0 &                                 10.4 ± 1.4 &   0.3068 ± 0.0013 \\
         3 &                   24 &                           0.0 ± 0.7 &                                    0.0 ± 1.4 &   0.3481 ± 0.0012 \\
           &                   27 &                          -2.4 ± 0.5 &                                  0.7 ± 1.5 &   0.3452 ± 0.0014 \\
\midrule
           &                 None &                         -15.2 ± 1.3 &                                     12 ± 4 &   0.3038 ± 0.0013 \\
         4 &                   24 &                             0.0 ± 0.7 &                                      0 ± 3 &   0.3498 ± 0.0016 \\
           &                   30 &                          -3.9 ± 1.3 &                                      0 ± 4 &   0.3463 ± 0.0012 \\
\midrule
           &                 None &                         -16.9 ± 1.5 &                                 13.1 ± 2.1 &   0.2984 ± 0.0012 \\
         5 &                   24 &                             0.0 ± 1.6 &                                    0.0 ± 2.1 &   0.3527 ± 0.0009 \\
           &                   36 &                          -6.4 ± 1.6 &                                  0.9 ± 1.8 &   0.3489 ± 0.0009 \\
\midrule
\bottomrule
\end{tabular}
\end{threeparttable}
\end{table}

\bibliographystyle{icml2014}

\end{document}